\documentclass[twocolumn,amsfonts,amssymb,amsmath,floatfix,final,letterpaper]{revtex4}

\usepackage{graphics}
\usepackage{verbatim}
\usepackage{amscd}
\usepackage{dcolumn}
\usepackage{bm}
\usepackage{longtable}

\newcommand{\lp}[1]{\left#1}
\newcommand{\rp}[1]{\right#1}

\newcommand{\unit}[1]{\,\mathrm{#1}}
\newcommand{\uAA}{\,\textrm{\AA}}
\newcommand{\eV}{\,\mathrm{eV}}
\newcommand{\meV}{\,\mathrm{meV}}
\newcommand{\mueV}{\,\mathrm{\mu{}eV}}
\newcommand{\icm}{\,\mathrm{cm^{-1}}}
\newcommand{\Hartree}{\,\mathrm{Hartree}}
\newcommand{\mHartree}{\,\mathrm{mHartree}}

\newcommand{\etal}{\emph{et al.}}
\newcommand{\ie}{\emph{i.e.}}
\newcommand{\eg}{\emph{e.g.}}
\newcommand{\vi}{\emph{vide infra}}
\newcommand{\adhoc}{\emph{ad hoc}}

\newcommand{\SSgp}{\mbox{${}^1\Sigma_g^+$}}
\newcommand{\SGg}{\mbox{${}^1\Gamma_g$}}
\newcommand{\SSum}{\mbox{${}^1\Sigma_u^-$}}
\newcommand{\TSup}{\mbox{${}^3\Sigma_u^+$}}
\newcommand{\TGu}{\mbox{${}^3\Gamma_u$}}
\newcommand{\TSgm}{\mbox{${}^3\Sigma_g^-$}}

\newcommand{\dxy}{d_{xy}}
\newcommand{\dxz}{d_{xz}}
\newcommand{\dyz}{d_{yz}}
\newcommand{\dzsq}{d_{z^2}}
\newcommand{\dxsqmysq}{d_{x^2-y^2}}

\newcommand{\vv}{\mbox{$(d_{xy}^Ad_{xy}^B)$}}
\newcommand{\uv}{\mbox{$(d_{x^2-y^2}^Ad_{xy}^B)$}}
\newcommand{\uu}{\mbox{$(d_{x^2-y^2}^Ad_{x^2-y^2}^B)$}}
\newcommand{\uSvv}{\mbox{${}^{1,3}(d_{xy}^Ad_{xy}^B)$}}
\newcommand{\uSuv}{\mbox{${}^{1,3}(d_{x^2-y^2}^Ad_{xy}^B)$}}
\newcommand{\uSuu}{\mbox{${}^{1,3}(d_{x^2-y^2}^Ad_{x^2-y^2}^B)$}}
\newcommand{\Svv}{\mbox{${}^1(d_{xy}^Ad_{xy}^B)$}}
\newcommand{\Suv}{\mbox{${}^1(d_{x^2-y^2}^Ad_{xy}^B)$}}
\newcommand{\Suu}{\mbox{${}^1(d_{x^2-y^2}^Ad_{x^2-y^2}^B)$}}
\newcommand{\Tvv}{\mbox{${}^3(d_{xy}^Ad_{xy}^B)$}}
\newcommand{\Tuv}{\mbox{${}^3(d_{x^2-y^2}^Ad_{xy}^B)$}}
\newcommand{\Tuu}{\mbox{${}^3(d_{x^2-y^2}^Ad_{x^2-y^2}^B)$}}

\newcommand{\ket}[1]{\vert #1\rangle}
\newcommand{\bra}[1]{\langle #1\vert}
\newcommand{\avg}[1]{\langle #1\rangle}

\newcommand{\NiHRC}{\textit{a}}
\newcommand{\NiDimerSORC}{\textit{a}}

\newcommand{\minh}{\rule{0pt}{2.5ex}}

\begin{document}

\title{Broken-Symmetry Unrestricted Hybrid Density Functional
       Calculations \\ on Nickel Dimer and Nickel Hydride}

\author{Cristian V. Diaconu}
\email{cvdiaconu@brown.edu}
\author{Art E. Cho}
	\altaffiliation{Present address: Department of Chemistry and Center
	for Biomolecular Simulation, Columbia University, New York, New
	York 10027.}
\author{J. D. Doll}
\affiliation{
Department of Chemistry, Box H, Brown University,\\
Providence, Rhode Island 02912
}

\author{David L. Freeman}
\affiliation{
Department of Chemistry, University of Rhode Island,\\
Kingston, Rhode Island 02881
}

\date{August 5, 2004}

\begin{abstract}
In the present work we investigate the adequacy of broken-symmetry
unrestricted density functional theory (DFT) for constructing the
potential energy curve of nickel dimer and nickel hydride, as a model
for larger bare and hydrogenated nickel cluster calculations. We use
three hybrid functionals: the popular B3LYP, Becke's newest optimized
functional Becke98, and the simple FSLYP functional ($50\,\%$
Hartree-Fock and $50\,\%$ Slater exchange and LYP gradient-corrected
correlation functional) with two basis sets: all-electron (AE)
Wachters+f basis set and Stuttgart RSC effective core potential (ECP)
and basis set.
We find that, overall, the best agreement with experiment, comparable
to that of the high-level CASPT2, is obtained with B3LYP/AE, closely
followed by Becke98/AE and Becke98/ECP. FSLYP/AE and B3LYP/ECP give
slightly worse agreement with experiment, and FSLYP/ECP is the only
method among the ones we studied that gives an unaceptably large
error, underestimating the dissociation energy of Ni$_2$ by $28\,\%$,
and being in the largest disagreement with the experiment and the
other theoretical predictions.
We also find that for Ni$_2$, the spin-projection for the
broken-symmetry unrestricted singlet states changes the ordering of
the states, but the splittings are less than $10\meV$.  All our
calculations predict a {\mbox{$\delta\delta$-hole}} ground state for
Ni$_2$ and {\mbox{$\delta$-hole}} ground state for NiH.  Upon
spin-projection of the singlet state of Ni$_2$, almost all of our
calculations: Becke98 and FSLYP both AE and ECP and B3LYP/AE predict
{\Suu{}} or {\Svv{}} ground state, which is a mixture of {\SSgp{}} and
{\SGg{}}. B3LYP/ECP predicts a {\Tuv{}} (mixture of {\TSgm{}} and
{\TGu{}}) ground state virtually degenerate with the {\Suu{}}/{\Svv{}}
state. The doublet {\mbox{$\delta$-hole}} ground state of NiH
predicted by all our calculations is in agreement with the
experimentally predicted $^2\Delta$ ground state.  For Ni$_2$, all our
results are consistent with the experimentally predicted ground state
of $0_g^+$ (a mixture of $\SSgp{}$ and $\TSgm{}$) or $0_u^-$ (a
mixture of $\SSum{}$ and $\TSup{}$).
\end{abstract}

\maketitle

\section{Introduction} \label{sec:intro}

During the last decades clusters have been extensively studied because
of their potential applications, their theoretical value in
understanding the transition from isolated atomic systems to condensed
matter~\cite{Castleman:86:rev, Castleman:96:rev} and their relevance
to the study of surface processes and heterogeneous
catalysis~\cite{Moskovits:91:rev, Pacchioni:92, Knickelbein:99:sp}.
The rapid development of experimental techniques in recent years has
made it possible both to obtain size-controlled transition metal
clusters and to study their reactivity against chemisorption
processes~\cite{Trevor:85, Zakin:87, Kaldor:88, Zakin:89}.

Methods for studying properties and behavior of clusters have been
developed, and a review on computational studies of clusters has been
written by Freeman and Doll~\cite{Freeman:96:rev}.  There have been
many studies on nickel clusters using various methods of exploring the
potential energy surfaces (PES).  The construction of such potential
surfaces is a major problem, especially for transition metal clusters.
Many methods have been used to construct PES's for nickel clusters,
ranging from empirical --- Finis-Sinclair type~\cite{Nayak:97,
Xiang:00}, semi-empirical --- tight binding~\cite{Lathiotakis:96,
Luo_C:00} and extended H{\"u}ckel~\cite{Curotto:98}, to \emph{ab
initio} or mixed empirical--\emph{ab initio}~\cite{Reddy:98}
approaches.  Recently, there have been studies of hydrogen atoms on Cu
surfaces~\cite{Bae:01:phd} within the density functional framework.
It has been found that semiempirical methods are insufficient for
accurate description of such systems, and first principle
quantum-mechanical methods are needed to obtain a proper description
of the hydrogen binding site.

Our long-term goal is to explore the structure and dynamics of
clusters, including nickel and nickel hydride systems. The combination
of the physical complexity and the computational demands of these
systems necessitate that the microscopic force laws that are utilized
in such simulations be both efficient and reliable.

Among the correlated electronic structure methods the best candidate
is clearly density functional theory (DFT) because of its ability of
reaching high accuracy --- similar to coupled-cluster CCSD(T) method
for second-row elements --- when hybrid exchange-correlation
functionals are used~\cite{Cremer:01}.  Moreover, DFT (using hybrid
functionals) is computationally not much more expensive than
Hartree-Fock.

While there have been a number of DFT calculations reported on small
nickel clusters~\cite{Basch:80, Tomonari:86, Mlynarski:91, Reuse:95,
Michelini:98, Michelini:99, Castro:97, Berces:97}, some results appear
to be inconsistent both with respect to available experimental data
and/or with respect to other theoretical predictions.

The works of Yanagisawa \emph{et al.}~\cite{YTH:00} and Barden
\emph{et al.}~\cite{Barden:00} on the performance of DFT on the first
transition metal series have shown that non-hybrid functionals
(BLYP~\cite{Becke:88,LYP}, BP86~\cite{Becke:88, Perdew:86:P86.1,
Perdew:86:P86.2}, BOP~\cite{Becke:88, Tsuneda:99:OP} and
PW91~\cite{PW:91}) and hybrid functionals (B3LYP~\cite{Becke:93:B3,
LYP}, BHLYP~\cite{Becke:93:BH, LYP}) give an overall similar
description for $3d$ transition metal dimers, with the non-hybrid ones
giving better bond lengths and the hybrid ones better dissociation
energies.  However, while Yanagisawa \emph{et al.}~\cite{YTH:00}
obtain good agreement with experiment for nickel dimer for all studied
exchange-correlation functionals (they only calculated the triplet
states), Barden \emph{et al.}~\cite{Barden:00} obtained a negative
dissociation energy for their calculated singlet ground state with the
B3LYP functional (and negative or very close to zero for all hybrid
exchange-correlation functionals).  This prompted us to use symmetry
breaking in unrestricted DFT for describing the lowest singlet state
of nickel dimer.  With larger cluster calculations in mind, we also
used broken symmetry unrestricted DFT to better describe bond breaking
in all states of nickel dimer and nickel hydride.

It has been argued that broken-symmetry unrestricted calculations
(Hartree-Fock and DFT with hybrid functionals) are useful for
describing systems with weakly coupled electron
pairs~\cite{Bernard:79, Noodleman:81, Fan:97, martin97:hybrid_dft,
illas98:fslyp, Grafenstein:01, Cremer:01}.  Ni$_2$ is definitely such
a case, as previously observed by Basch \etal{}~\cite{Basch:80}.  As
argued by Cremer~\cite{Cremer:01}, the combination of hybrid
exchange-correlation functional with symmetry breaking leads to a
better description of systems in which static correlation is present
than does the restricted DFT formalism.  Finally, we believe that the
formalism used to describe any system is solely dictated by the
objective of the calculation.  For a variational approach, and DFT can
be regarded as such --- aside from the exchange-correlation
functional, --- the more flexible is the form of the trial function
(density), the lower is the obtained energy.  Since our interest is
mainly in the energetics of nickel clusters, the best choice for us
seems to be the unrestricted broken symmetry DFT approach with hybrid
functionals.

In the present work we study the nickel dimer and nickel hydride using
broken symmetry unrestricted DFT with hybrid exchange-correlation
functionals --- mainly the popular B3LYP~\cite{Becke:93:B3, LYP} ---
as model systems for larger bare and hydrogenated nickel clusters in
an attempt to establish what might comprise a minimally reliable
method for more extensive nickel cluster calculations.

The outline of the reminder of the present paper is as follows:
Section~\ref{sec:methods} contains discussions the methods used,
Section~\ref{sec:results} presents the results of the calculations,
and, where possible, comparisons with previous reports.
Section~\ref{sec:conclusions} concludes with suggestions for further
research based on the present findings.


\section{Computational details} \label{sec:methods}

The DFT calculations reported in this paper are carried out with
NWChem~\cite{NWChem} computational chemistry package, using the
unrestricted Kohn-Sham approach, allowing for symmetry breaking, and
using a finite orbital (spherical Gaussian) basis set expansion and
charge density fitting.

Hartree-Fock (HF) and second order M{\o}ller-Plesset (MP2)
calculations are performed for comparison for the states of the Ni
atom and are done in unrestricted form.\footnote{For the closed shell
${}^1S$ state of Ni, both restricted and unrestricted HF and MP2
calculations are performed and the difference between the restricted
and the unrestricted energies is of the same order of magnitude as the
convergence criterion ($10^{-8}\Hartree$).}

Throughout the paper we will use the notation $^{M}(h^Ah^B)$ for the
states of nickel dimer, where $M$ is the multiplicity, $h^A$ and $h^B$
are the unoccupied (hole) orbitals in the $3d$ shell on the two Ni
atoms, denoted $A$ and $B$.  The broken symmetry singlet states (with
$S_z=0$ and $\avg{S^2}=1$) are denoted by $^{1,3}(h^Ah^B)$.

In general, an unrestricted Slater or Kohn-Sham determinant is not an
eigenfunction of the total spin operator $S^2$, and the results can
only be characterized by the number of $\alpha$ and $\beta$ electrons.
However, following common usage, we refer to the states that differ in
the number of $\alpha$ and $\beta$ electrons by 0 as singlets, by 1 as
doublets, and so on.  We explicitly identify pure spin states where
relevant.


\subsection{Exchange-Correlation Functionals}

We used three hybrid exchange-correlation functionals: the very
popular B3LYP --- composed of the B3, Becke's three-parameter hybrid
exchange functional~\cite{Becke:93:B3} and LYP~\cite{LYP} correlation
functional --- is the first choice because it is well known and
extensively characterized.  Becke's newest optimized functional,
Becke98~\cite{becke98} is also used, since it is supposed to be, in a
certain sense, the best obtainable exchange-correlation functional
within the gradient-corrected framework.  The hybrid composed of half
Slater exchange~\cite{Slater:74:X_alpha}, half Hartree-Fock exchange
and LYP~\cite{LYP} correlation, named here FSLYP is also used for
comparison, as it is the simplest theoretically-justifiable hybrid
method and is reported to perform rather well~\cite{illas98:fslyp,
martin97:hybrid_dft}.


\subsection{Basis sets}

All calculations are performed with spherical basis sets.  As
all-electron (AE) basis sets, Wachters+f basis
set~\cite{basis:Wachters.1, basis:Wachters.2, basis:3dTM-Hay-aug,
basis:Wachters+f}, a $[14s11p6d3f]$/$(8s6p4d1f)$ contraction is used
for nickel and 6-311++G(2d,2p), a $[6s2p]$/$(4s2p)$ contraction for
hydrogen.

Effective-core potentials (ECP) are also explored, since they greatly
reduce computational cost.  Stuttgart RSC ECP effective core
potentials basis
set~\cite{stuttgart_rsc_ecp:87,stuttgart_rsc_ecp:87:note}
are used for nickel, as they provide a similar quality of valence
basis functions as Wachters+f.

Ahlrichs Coulomb Fitting~\cite{Ahlrichs_cd:95, Ahlrichs_cd:97} basis
is used as a charge density (CD) fitting basis only for the
all-electron calculations, as it significantly reduces computing time,
especially for larger systems. When not specified otherwise, all reported
all-electron results are obtained using charge density fitting. 

We did not use charge density fitting with ECP because of the large
errors that resulted when we tried the use of Ahlrichs Coulomb Fitting
basis in combination with Stuttgart RSC ECP. For example, for B3LYP
functional, CD fitting error is as much as $0.3\eV$ for both the
interconfigurational energies of Ni atom and the binding energy of
Ni$_2$.  Please refer to Appendix~\ref{sec:Accur:CDfit} for discussion
of the accuracy of charge density fitting.


\subsection{Numerical integration and convergence}

The numerical integration necessary for the evaluation of the
exchange-correlation energy implemented in NWChem uses an
Euler-MacLaurin scheme for the radial components (with a modified
Mura-Knowles transformation) and a Lebedev scheme for the angular
components.  We use three levels of accuracy for the numerical
integration that are used in our DFT calculations, labeled by the
corresponding keywords from NWChem ({\verb|medium|}, {\verb|fine|} and
{\verb|xfine|}).

The reported atomic calculations are those obtained with the
{\verb|xfine|} grid.  For geometry optimization and vibrational
frequency calculations we use the {\verb|fine|} grid.  And for the
potential energy curve (PEC) scans we used the {\verb|medium|} grid.
The maximum number of iterations is set to 100 in all calculations.

Please refer to Appendix~\ref{sec:Accur:DFT} for details on numerical
integration and convergence criteria.


\subsection{Initial guess}

For all DFT methods we first performed a calculation for Ni atom using
fractional occupation numbers (FONs)~\cite{Warren:1996:FONs}, as
implemented in NWChem. We use an exponent of $0.01\Hartree$ for the
Gaussian broadening function. We then use the molecular orbitals from
the FONs calculation, after proper reordering, as initial guess for
computing the $^3F$ and $^3D$ states of Ni atom. We use
$t_{2g}^6e_g^2$ configuration in the $O_h$ symmetry group for the
$^3F$ state. In order to obtain the lowest energy possible for the
$^3D$ state, we scan all hole positions: $\dzsq$, $\dxsqmysq$ and
$\dxy$ using $D_{4h}$ symmetry group, and $\dxz$ and $\dyz$ using
$D_{2h}$ symmetry group, enforcing the position of the hole with a
maximum overlap condition.

For Ni$_2$ and NiH, we use a broken-symmetry initial guess of the
form: $3d^94s^1\uparrow\uparrow+\downarrow\downarrow3d^94s^1$ for
singlet Ni$_2$, $3d^94s^1\uparrow\uparrow+\downarrow\uparrow3d^94s^1$
for triplet Ni$_2$ and
$\text{Ni}\,3d^94s^1\uparrow\uparrow+\downarrow1s^1\,\text{H}$ for
NiH. As initial guess molecular orbitals we use those from the Ni atom
calculations, sweeping through all unique positions of the holes in
the $3d$ orbitals of Ni atom(s), and enforcing the position of the
hole(s) with a maximum overlap condition.


\subsection{Geometry optimization}

Geometry optimizations are performed using the DRIVER module of NWChem
using NWChem's default convergence criteria (in atomic units):
$4.5\cdot{}10^{-4}$ maximum and $3.0\cdot{}10^{-4}$ root mean square
gradient, $1.8\cdot{}10^{-3}$ maximum and $1.2\cdot{}10^{-3}$ root
mean square of the cartesian step.  These convergence criteria give a
maximum error in equilibrium bond length of less than
$\approx{}10^{-3}\uAA$ for Ni$_2$ and less than
$\approx{}5\cdot{}10^{-4}\uAA$ for NiH.  The available precision is set
to $5\cdot{}10^{-7}\Hartree$ for the \verb|fine| grid and
$5\cdot{}10^{-8}\Hartree$ for the \verb|xfine| grid.

\subsection{Vibrational frequencies}

Harmonic vibrational frequencies are calculated using NWChem's VIB
module with the default options.  Since analytical Hessian for open
shell systems is not available for the exchange-correlation
functionals used, the Hessian is computed by finite differences with
$\Delta=0.01\unit{Bohr}$, which gives an estimated error for the
vibrational frequencies of $\approx{}0.5\icm$ ($\approx{}0.25\,\%$)
for Ni$_2$ and $\approx{}2\icm$ ($\approx{}0.1\,\%$) for NiH.


\subsection{Spin and Symmetry Projection}

In general, an open-shell Slater or Kohn-Sham determinant is not an
eigenfunction of the total spin operator $S^2$.  However, spin-adapted
configurations can be obtained as combinations of (a small number of)
restricted determinants \cite{Szabo:qc_text, paunz79:spin_eigenfns}.
Unrestricted determinants are not eigenfunction of the total spin
operator $S^2$, either, and they cannot be spin-adapted by combining a
small number of unrestricted determinants~\cite{Szabo:qc_text}.
However, for antiferromagnetic coupling of two weakly interacting
identical high spin monomers, Noodleman~\cite{Noodleman:81} derived an
approximate spin projection scheme that is correct to the first order
in the overlap integrals.  Ni$_2$ can be well approximated by such a
model.

As previously observed by Basch \etal{}~\cite{Basch:80}, the
electronic structure of nickel clusters corresponds roughly to a model
in which the $3d$ electrons can be viewed as weakly interacting
localized $3d^9$ units bound together primarily by $4s$ electrons.  If
the $4s$ electrons are paired in a $\sigma$ bond, then Ni$_2$ has two
possible spin states: singlet and triplet.  However, the open-shell
singlet state can not be represented by a single determinant, and the
broken-symmetry single determinant $\Psi_B$ obtained by putting one of
the open-shell electrons in a spin $\alpha$ $d$ orbital on one of the
Ni atoms and the other electron in a spin $\beta$ $d$ orbital on the
other Ni atom is not pure singlet, but an equal mixture of singlet and
triplet (using $\ket{S,S_z}$ notation for the spin states):
\[
   \Psi_B=\frac{1}{\sqrt2}\ket{0,0}+\frac{1}{\sqrt2}\ket{1,0}
\]
with the expectation value of the total spin
$\bra{\Psi_B}{S^2}\ket{\Psi_B}=1$.  In agreement with this model, for
the broken symmetry calculations of the $S_z=0$ state of the Ni dimer
the expectation value of the total spin $\avg{S^2}$ is close to the
exact value of $1$ for the broken-symmetry mixed state, and for the
triplet ($S_z=1$) state, $\avg{S^2}$ is close to the exact value of
$2$ (in both cases, the relative absolute differences between the
computed and the exact values are less than $2\,\%$).  Mulliken
population analysis also supports the weakly interacting $3d^9$ units
model.  For the triplet nickel dimer there is a Mulliken spin
population of 1.00 on each Ni atom, and for the broken symmetry
singlet there is a Mulliken spin population of $1.1$ on one of the Ni
atoms and $-1.1$ on the other.

Using the approximate projection method of Noodleman~\cite{Noodleman:81},
the energy of the pure singlet state, $E(0)$ can be obtained from the
energy of the unrestricted broken-symmetry singlet,~$E_B$, and the
energy of the triplet,~$E(1)$:
\begin{equation} \label{eq:spin_proj_nosym}
   E(0)=2E_B-E(1).
\end{equation}
 The same result can be also obtained by the
spin projection technique (see, e.g., Refs.~\onlinecite{Daul:2003} and
\onlinecite{Cramer:2002:EssenComputChem}).

Ni$_2$ belongs to $D_{\infty h}$ point symmetry group, and the
irreducible representations (irreps.) are good quantum numbers for the
molecular states.  We combine the spin projection with symmetry
projection to extract the maximum information possible from the
single-determinant Kohn-Sham DFT calculations.  From simple
group-theoretical considerations one can find that the pure spin and
symmetry states of Ni$_2$ that arise from $d_\delta$ orbitals, which
are found to give the lowest energy states for all calculations, are:
$^1\Sigma_g^+$, $^1\Gamma_g$, $^1\Sigma_u^-$, $^3\Sigma_g^-$,
$^3\Sigma_u^+$ and $^3\Gamma_u$.
Within the model of two weakly interacting $3d^9$ units, for the
purpose of projection we consider only the active electrons and the
active orbitals on each center, namely $d_{x^2-y^2}^A$, $d_{xy}^A$,
$d_{x^2-y^2}^B$ and $d_{xy}^B$.

The projection has been carried out using the projection operators
technique in $D_{8h}$, which the smallest subgroup of $D_{\infty h}$
in which all irreps. arising from the $(d_\delta^A)^1(d_\delta^B)^1$
configuration can be completely correlated, and the following
equations relating the energies of the pure spin and symmetry states
listed above to the energies of the computed triplet and projected
singlet states are obtained:

\begin{subequations} \label{eq:sym_proj}
\begin{align}
   E\lp({\Suu}\rp) & = \frac{1}{2}\lp[ E\lp(^1\Sigma_g^+\rp)
                                      +E\lp(^1\Gamma_g\rp)\rp]  \\
   E\lp({\Tuu}\rp) & = \frac{1}{2}\lp[ E\lp(^3\Sigma_u^+\rp)
                                      +E\lp(^3\Gamma_u\rp) \rp] \\
   E\lp({\Svv}\rp) & = \frac{1}{2}\lp[ E\lp(^1\Sigma_g^+\rp)
                                      +E\lp(^1\Gamma_g\rp)\rp]  \\ 
   E\lp({\Tvv}\rp) & = \frac{1}{2}\lp[ E\lp(^3\Sigma_u^+\rp)
                                      +E\lp(^3\Gamma_u\rp) \rp] \\ 
   E\lp({\Suv}\rp) & = \frac{1}{2}\lp[ E\lp(^1\Sigma_u^-\rp)
                                      +E\lp(^1\Gamma_g\rp)\rp]  \\
   E\lp({\Tuv}\rp) & = \frac{1}{2}\lp[ E\lp(^3\Sigma_g^-\rp)
                                        +E\lp(^3\Gamma_u\rp)\rp].
\end{align}
\end{subequations}

These equations contain the maximal information that can be obtained
from single-determinant calculations.  

From equations~(\ref{eq:sym_proj}) we can derive the (partially)
symmetry-adapted equivalent of Eq. (\ref{eq:spin_proj_nosym}):
\begin{equation} \label{eq:spin_proj_sym}
   E\lp(^{1}(h^Ah^B)\rp) = 2E\lp(^{1,3}(h^Ah^B)\rp) - E\lp(^{3}(h^Ah^B)\rp)
\end{equation}
where $(h^Ah^B)$ represents each of {\uu{}}, {\uv{}}, and {\vv{}}.
The spin projection has to be done separately for each of the
combinations of holes {\uu{}} and {\uv{}}.

Since the equations for the states $^M\vv$ have a similar form to
those for the $^M\uu$ states, $^M\uu$ and $^M\vv$ states should have
the same energy ($M$ can be $1$, $3$ or $(1,3)$).  We calculate the
{\uSvv{}} and {\Tvv{}} states for consistency check.

Since the bond lengths for the pure spin states are different from
each other and from the mixed state, we use a harmonic approximation
of the potential around equilibrium bond length for each state:
\[ 
   E(d) = -D_e + \frac{1}{2}\mu\omega_e^2(d-d_e)^2
\]
and solve the resulting equations for $d_e$ (equilibrium bond lenth),
$D_e$ (dissociation energy) and $\omega_e$ (vibrational frequency) for
the projected state (here $\mu$ denotes the reduced mass of the
molecule).


\section{Results and Discussions} \label{sec:results}

\subsection{Nickel Atom} \label{ssec:atom}

%
\begin{table*}[tbp]
\caption[Atomic states of Ni]{\label{tab:Ni-atom}
Energies of atomic states of Ni.  Values are in eV, relative to the
ground state.}
\begin{minipage}{\textwidth}
\begin{ruledtabular}
\begin{tabular}{lddddddddddd}
 &
 &
 &
 & \multicolumn{1}{c}{UHF} 
 & \multicolumn{1}{c}{MP2} 
 & \multicolumn{2}{c}{FSLYP} 
 & \multicolumn{2}{c}{B3LYP} 
 & \multicolumn{2}{c}{Becke98} 
\\
\cline{7-8}\cline{9-10}\cline{11-12}
State
 & \multicolumn{1}{c}{Exp.}\footnote{Weighted averages over the $J$
components of the experimental values~\cite{Sugar:85}.}
 & \multicolumn{1}{c}{RC}\footnote{Martin and Hay estimations of
relativistic corrections from Ref.~\mbox{\onlinecite{Martin+Hay:81}}.}
 & \multicolumn{1}{c}{Exp.-RC}\footnote{Experimental values with
relativistic corrections subtracted.}
 & \multicolumn{1}{c}{AE} 
 & \multicolumn{1}{c}{AE} 
 & \multicolumn{1}{c}{AE} 
 & \multicolumn{1}{c}{ECP} 
 & \multicolumn{1}{c}{AE} 
 & \multicolumn{1}{c}{ECP} 
 & \multicolumn{1}{c}{AE} 
 & \multicolumn{1}{c}{ECP}\minh
\\
\hline
$^3D(3d^94s^1)$  & 0    &       & 0      & 1.44 & 0.27 & 0.12  & 0.32  & 0     & 0.01  & 0    & 0.20\minh \\
$^3F(3d^84s^2)$  & 0.03 & -0.36 & 0.39   & 0    & 1.41 & 0     & 0     & 0.36  & 0     & 0.29 & 0    \\
$^1S(3d^{10})$   & 1.74 &  0.21 & 1.53   & 5.81 & 0    & 2.62  & 3.02  & 1.90  & 2.21  & 1.78 & 2.37 
\end{tabular}
\end{ruledtabular}
\end{minipage}
\end{table*}

The ground state of the nickel atom is
$^3F_4(3d^84s^2)$~\cite{Moore:52,Sugar:85}.  However, since our
calculations do not include spin-orbit coupling, we use weighted
averages over the $J$ components of the experimental data for
comparison, which makes $^3D(3d^94s^1)$ the ground state, with
$^3F(3d^84s^2)$ state only $0.03\unit{eV}$ higher, and $^1S(3d^{10})$
state $1.74\unit{eV}$ above the ground state.

As first estimated by Martin and Hay~\cite{Martin+Hay:81} and
confirmed by full relativistic calculations done by Jeng and
Hsue~\cite{Jeng+Hsue:00} the relativistic effects in the $3d$
transition metal series are important.  Therefore, in comparing our
non-relativistic calculations with the experiment we take such effects
into account by subtracting the estimated values reported by Martin
and Hay from the experimental values.  After this correction (see
Table~\ref{tab:Ni-atom} for details), the ground state remains $^3D$,
with $^3F$ state $0.39\unit{eV}$ higher, and $^1S$ state
$1.53\unit{eV}$ above the ground state.  These values will be referred
to as ``relativistically corrected (RC) experimental values.''

In Table~\ref{tab:Ni-atom} we choose to utilize the Martin and
Hay~\cite{Martin+Hay:81} relativistic corrections as opposed to the
ones computed by Jeng and Hsue~\cite{Jeng+Hsue:00} because they
include the additional ${}^1S(3d^{10})$ configuration.  The results of
the recent relativistic calculations in the RESC approximation
(relativistic scheme by eliminating small components) reported by
Yanagisawa \etal~\cite{YTH:00} do not lend themselves to an analysis
of relativistic corrections.  Moreover, these calculations seem to be
at odds with the two previous calculations.

Our results, summarized in Table~\ref{tab:Ni-atom}, show that only the
DFT/Wachters+f calculations with B3LYP and Becke98 hybrid
exchange-correlation functionals predict a $^3D$ ground state,
although B3LYP/ECP predicts the $^3D$ state only $0.01\eV$ above the
$^3F$ ground state.

It is worth mentioning that, for all our DFT calculations, there are
differences between the components of the $^3D$ state of Ni and these
differences range from $4\meV$ to $37\meV$. We report the energy of
the $^3D$ component with the lowest energy as the energy of the $^3D$
state. It is also worth mentioning that the B3LYP/ECP calculations
fail to converge for the spin $\alpha$ $\dxy$-, $\dyz$-, $\dxz$-, and
$\dxsqmysq$-hole components of the $^3D$ state.

The all-electron calculations with B3LYP and Becke98 XC functionals
also predict an ordering of the $^3D$, $^3F$ and $^1S$ states in
agreement with the experiment. 

The values of the computed energies of $^3F$ (relative to $^3D$)
differ from the observed experimental values by $0.30\eV$ (B3LYP) and
$0.26\eV$ (Becke98). However, when compared with the relativistically
corrected experimental values, the differences drop to only $-0.06\eV$
and $-0.10\eV$, respectively. On the other hand, the computed energies
of $^1S$ (relative to $^3D$) are larger than the observed experimental
values by $0.16\eV$ (B3LYP) and $0.04\eV$ (Becke98), and larger than
the relativistically corrected experimental values by $0.37\eV$ and
$0.25\eV$, respectively. However, the larger errors in the $^1S$ is
less important for the purpose of nickel cluster calculations.

Hartree-Fock calculations predict $^3F$ ground state, $^3D$
$1.44\unit{eV}$ higher and ${}^1S$ $5.81\eV$ above the ground state in
good agreement with numerical HF calculations of Martin and
Hay~\cite{Martin+Hay:81}, but with large errors compared to the RC
experimental values.  MP2 calculations predict $^1S$ ground state,
with $^3D$ and $^3F$ states $0.27\unit{eV}$ and $1.41\unit{eV}$
higher, respectively.

The unoptimized FSLYP functional is, as expected, the least
accurate. With the Wachters+f basis it yields results that differ from
the RC experimental values and B3LYP and Becke98 results by $\approx
-0.5\eV$ for $^3F$ and by $\approx 0.5\eV$ for $^1S$.

The effective core potentials (ECP) tend to overstabilize $^3F$ by
$0.2-0.5\eV$ and destabilize $^1S$ by $0.2-0.4\eV$ (relative to $^3D$)
with respect to the all-electron counterparts. Thus, all our DFT/ECP
calculations predict $^3F$ ground state. However, the B3LYP/ECP
calculations yield $^3D$ only $0.01\eV$ above the $^3F$ ground state,
which can be considered acceptable error for the dissociation energy
of nickel dimer which is of order of $2\eV$, given the savings of
using ECPs.


\subsection{Nickel Dimer} \label{ssec:dimer}

%
\begin{table*}[tbp]
\caption[Ground state of Ni$_2$ -- comparison between computations and
experiment.]{\label{tab:Ni2-gs}
Ground state of Ni$_2$ -- comparison between computations and
experiment.  The reported singlet states from our calculations are
projected.  $d_e$ -- bond length ($\uAA$), $D_e$ -- dissociation
energy, relative to ground state Ni atoms (without zero-point
correction, $\eV$), $\omega_e$ -- vibrational frequency
($\unit{cm}^{-1}$).  The relative deviations from the experimental
values are given in parentheses, and the average (AARD) and maximum
(MARD) absolute relative deviations from experimental values of $d_e$,
$D_e$ and $\omega_e$ are listed under AARD and MARD columns,
respectively.}
\begin{minipage}{\textwidth}
\begin{ruledtabular}
\begin{tabular}{ll@{\hspace*{1.0em}}l@{}r@{\hspace*{1.0em}}l@{}r@{\hspace*{1.0em}}l@{}r@{\hspace*{1.0em}}d@{\hspace*{1.0em}}d}
 Method
 & State
 & \multicolumn{2}{c@{\hspace*{1.0em}}}{\mbox{$d_e$}}
 & \multicolumn{2}{c@{\hspace*{1.0em}}}{\mbox{$D_e$}}
 & \multicolumn{2}{c@{\hspace*{1.0em}}}{\mbox{$\omega_e$}}
 & \multicolumn{1}{c@{\hspace*{1.0em}}}{\mbox{AARD}}
 & \multicolumn{1}{c}{\mbox{MARD}}
\\
\hline
FSLYP/ECP      & \Svv          & 2.236 & (1.5) & 1.325 & (-28.4) & 283.0 & (14.9) & 14.9 & 28.4\minh \\
FSLYP/AE       & \Svv          & 2.260 & (2.5) & 1.664 & (-10.1) & 271.1 & (10.1) &  7.6 & 10.1 \\
Becke98/AE     & \Svv          & 2.296 & (4.2) & 2.071 &  (11.9) & 256.8 &  (4.3) &  6.8 & 11.9 \\
CASPT2\footnote{We report here the values from Table VIII of
Ref.~\mbox{\onlinecite{Pou-Amerigo:94}}, last column ($+3s3p$ for
$d_e$ and $\omega_e$, and BSSE for $D_e$), from which we subtract the
estimated relativistic corrections (RC) and, for $D_e$ only, the
estimated spin-orbit coupling contributions (SO). From the same table
we estimate the relativistic corrections to $d_e$, $D_e$ and
$\omega_e$ as the difference between the values in the +RC column and
ones in the CASSCF column, and the spin-orbit coupling contribution to
$D_e$ as the difference between the value in the +SO column and the
one in the $+3s3p$ column. We also subtract these RC and SO
contributions from the experimental values.}
               & \SSgp,\SGg    & 2.281 & (3.5) &  1.89 &   (2.2) & 281.0 & (14.1) &  6.6 & 14.1 \\
CASSCF/IC-ACPF\footnote{from Ref.~\onlinecite{Bauschlicher:92}}
               & $^1\Gamma_g$  & 2.291 & (3.9) & 1.691 &  (-8.6) & 253.0 &  (2.8) &  5.1 &  8.6 \\
Becke98/ECP    & \Suu          & 2.278 & (3.4) & 1.792 &  (-3.1) & 265.1 &  (7.7) &  4.7 &  7.7 \\
B3LYP/ECP      & \Tuv          & 2.271 & (3.0) & 1.851 &   (0.1) & 269.3 &  (9.4) &  4.2 &  9.4 \\
B3LYP/AE       & \Suu          & 2.291 & (3.9) & 1.835 &  (-0.8) & 258.9 &  (5.2) &  3.3 &  5.2 \\
Exp.\footnote{Experimental values from which we subtract the CASPT2
estimates (see footnote~\NiDimerSORC) from
Ref.~\mbox{\onlinecite{Pou-Amerigo:94}} for relativistic contributions
(RC) for $d_e$, $D_e$ and $\omega_e$, and spin-orbit contributions
(SO) for $D_e$.  The experimental value of $d_e$ is
$2.1545\pm{}0.0004\uAA$~\cite{Morse:95} from which we subtract the
CASPT2 RC of $-0.05\uAA$.  The experimental value of $D_0$ is
$2.042\pm{}0.002\eV$~\cite{Morse:95}, from which we subtract the
CASPT2 RC of $0.07\eV$ and CASPT2 SO of $0.14\eV$; we report
$D_e=D_0+\frac{1}{2}\hbar\omega_e$. The experimental value of
$\omega_e$ is $259.2\pm 3.0\unit{cm}^{-1}$~\cite{Wang:96} from which
we subtract the CASPT2 RC of $13\unit{cm}^{-1}$. An earlier
work~\cite{Ho:93} reported $280\pm 20\unit{cm}^{-1}$.}
               & $0_g^+/0_u^-$ & 2.204 &       &  1.85 &         & 246.2 &        &      &      \\
\end{tabular}
\end{ruledtabular}
\end{minipage}
\end{table*}

The determination of the ground state of Ni$_2$ has been debated over
the last few decades. According to the recent
results~\cite{Pou-Amerigo:94,Morse:95}, the most plausible candidates
are spin-orbit coupled states of $\Omega=0_g^+$ (a mixture of
$^3\Sigma_g^-$ and $^1\Sigma_g^+$) and $\Omega=0_u^-$ (a mixture of
$^3\Sigma_u^+$ and $^1\Sigma_u^-$).

The bond lengths ($d_e$), dissociation energies ($D_e$)\footnote{All
dissociation energies reported are calculated relative to the energy
of the ground state of Ni atom(s) (without zero-point correction),
unless otherwise specified.} and vibrational frequencies ($\omega_e$)
for the ground state of Ni$_2$ from different calculations are
reported in Table~\ref{tab:Ni2-gs} along with experimental values and
results from other theoretical studies. The results in
Table~\ref{tab:Ni2-gs} are listed in the order of decreasing average
absolute relative deviations (AARD) from experimental values of bond
length ($d_e$), dissociation energy ($D_e$) and vibrational frequency
($\omega_e$).

Please note that our calculations are non-relativistic and do not
include spin-orbit coupling, and spin-orbit deperturbed values of
molecular properties of interest for Ni$_2$ are not available in the
literature. To account for that, we have subtracted the CASPT2
relativistic corrections (RC) to $d_e$, $D_e$ and $\omega_e$, and
spin-orbit contributions (SO) to $D_e$ from the experimental
values. We estimate the relativistic and spin-orbit coupling
corrections from Ref.~\mbox{\onlinecite{Pou-Amerigo:94}}. Please see
footnote~\NiDimerSORC\ of Table~\ref{tab:Ni2-gs} for details.

The reported singlet states from our calculations are spin-projected
by the approximate method described in Section~\ref{sec:methods}
(Computational details).

For the results from FSLYP/ECP and Becke98/ECP computations, the
splitting between the $(d_{xy}^Ad_{xy}^B)$ and
$(d_{x^2-y^2}^Ad_{x^2-y^2}^B)$ states, both for triplet and for mixed
$S_z=0$, is larger ($8\meV$ for FSLYP/ECP and $4\meV$ for Becke98/ECP)
than the accuracy of the DFT calculations (better than $0.1\meV$).
Thus, our approximate spin and symmetry projections are questionable
for these particular calculations. However, since we observed even
larger differences between the components of the $^3D$ state of Ni (up
to $0.03\eV$), we chose not to investigate this matter any further. In
these cases, the reported values are those of the component with the
lowest total energy (largest dissociation energy).

Almost all of our calculations: Becke98 and FSLYP both AE and ECP and
B3LYP/AE predict {\Suu{}} or {\Svv{}} ground state, which is a mixture of
{\SSgp{}} and {\SGg{}}. B3LYP/ECP predicts a {\Tuv{}} (mixture of
{\TSgm{}} and {\TGu{}}) ground state virtually degenerate with the
{\Suu{}}/{\Svv{}} state, which is only $1\meV$ higher in energy than
{\Tuv{}} ground state.

Among the high-level wavefunction methods, CASPT2 without spin-orbit
coupling predicts {\SSgp{}} ground state degenerate with {\SGg{}}, and
CASSCF/IC-ACPF predicts {\SGg{}} ground state.  Our DFT all-electron
calculations can be consistent with either one of the wavefunction
methods.  The experimental results are consistent with any of the
predictions of our DFT calculations and CASPT2, but not with the
{\SGg{}} state predicted by CASSCF/IC-ACPF.

\begin{figure}[tbp]
	\resizebox{3.375in}{!}{\includegraphics*{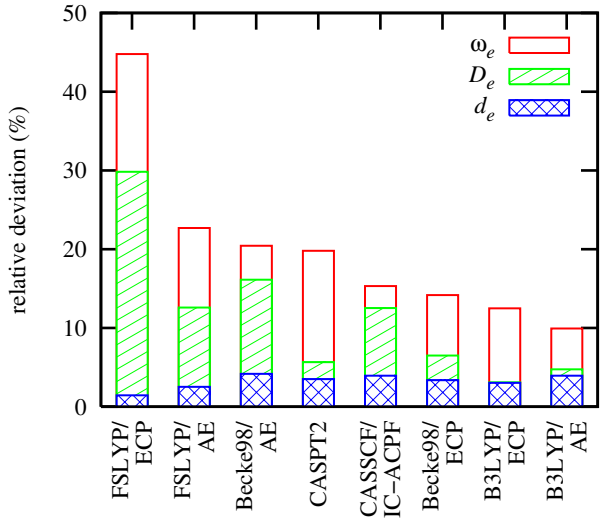}}%
\caption[Ni$_2$: performance of the used methods]%
{\label{fig:Ni2-gs-ARD} The absolute relative deviations from
experiment of computed dissociation energy, bond length, and
vibrational frequency for Ni$_2$. The results are arranged from left
to right in order of decreasing total absolute relative deviation
(TARD) --- the sum of absolute relative deviations from the
experimental values of the computed $d_e$, $D_e$ and $\omega_e$.}
\end{figure}

The absolute relative deviations from the experimental values of
computed bond lengths ($d_e$), dissociation energies ($D_e$) and
vibrational frequencies ($\omega_e$) for Ni$_2$ are plotted in
Fig.~\ref{fig:Ni2-gs-ARD}, arranged from left to right in order of
decreasing total absolute relative deviation (TARD) --- the sum of
absolute relative deviations from the experimental values of the
computed $d_e$, $D_e$ and $\omega_e$. 

From Fig.~\ref{fig:Ni2-gs-ARD}, as well as from
Table~\ref{tab:Ni2-gs}, it is apparent that overall, for Ni$_2$ the
all-electron DFT calculations with B3LYP functional give the best
agreement with experiment ($9.9\,\%$ TARD). B3LYP/ECP ($12.5\,\%$
TARD) and Becke98/ECP ($14.2\,\%$ TARD) follow with an overall
performance just a little better than CASSCF/IC-ACPF ($15.3\,\%$
TARD). Becke98/AE ($20.4\,\%$ TARD) and FSLYP/AE ($22.7\,\%$ TARD) are
next among our DFT calculations, performing just a few percent worse
than CASPT2 ($19.8\,\%$ TARD). With $44.8\,\%$ TARD, the FSLYP/ECP
calculation gives the largest disagreement with experiment and the
other methods.

%

The relative deviations from the experimental values of the computed
bond length ($d_e$), dissociation energy ($D_e$), asymptotic
dissociation energy ($D_e^a$, \vi) and vibrational frequency
($\omega_e$) of Ni$_2$ are plotted in Fig.~\ref{fig:Ni2-gs-RD} for
comparison. The values are arranged in order of increasing deviation
in the bond length.

\subsubsection{Bond length} \label{sssec:bondlength}

\begin{figure}[tbp]
	\resizebox{3.375in}{!}{\includegraphics*{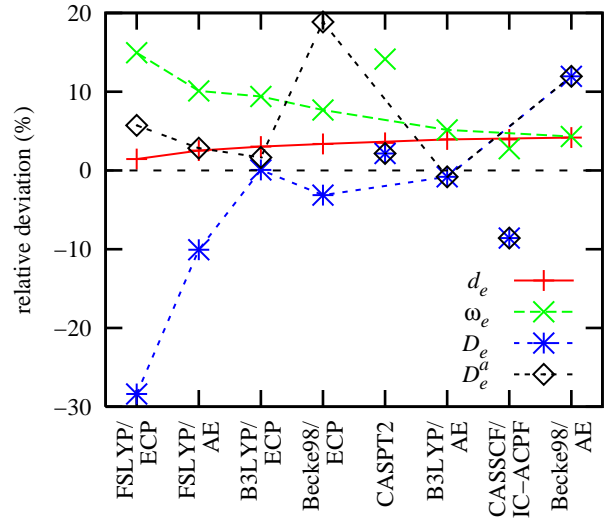}}%
\caption[Bond Length errors for Ni$_2$]%
{\label{fig:Ni2-gs-RD} The relative deviations from experimental
values of the computed bond length ($d_e$), dissociation energy
($D_e$), asymptotic dissociation energy ($D_e^a$, see text for
definition) and vibrational frequency ($\omega_e$) of Ni$_2$. Only the
results from our DFT calculations are connected by lines. CASPT2 and
CASSCF/IC-ACPF are included for comparison.}
\end{figure}

It is apparent that all calculations included in
Table~\ref{tab:Ni2-gs} and Fig.~\ref{fig:Ni2-gs-RD} --- both our
DFT calculations and the CASPT2~\cite{Pou-Amerigo:94} and
CASSCF/IAACPF~\cite{Bauschlicher:92} wavefunction methods included for
comparison --- overestimate the bond length of Ni$_2$. The deviations
from the experimental value of the computed bond length, $\Delta d_e =
d_e^\text{comp}-d_e^\text{exp}$ range between $0.03\uAA$ ($1.5\,\%$)
and $0.09\uAA$ ($4.2\,\%$).

Among our DFT calculations, the best agreement with the experiment for
the bond length of Ni$_2$ is obtained by FSLYP/ECP with $\Delta d_e =
0.032\uAA$ ($1.5\,\%$), followed by FSLYP/AE with $\Delta d_e =
0.056\uAA$ $(2.5\,\%)$ and B3LYP/ECP with $\Delta d_e = 0.067\uAA$
$(3.0\,\%)$. Becke98/ECP with $\Delta d_e = 0.074\uAA$ $(3.4\,\%)$
performs very similar to CASPT2, for which $\Delta d_e = 0.077\uAA$
$(3.5\,\%)$. Both B3LYP/AE and CASSCF/IC-ACPF are among the methods
that give the largest disagreement with the experiment, with $\Delta
d_e = 0.087\uAA$ $(3.9\,\%)$. Finally, Becke98/AE yields the worst
deviation from experiment, $\Delta d_e = 0.092\uAA$ $(4.2\,\%)$.

%

Among all three XC functionals, the best agreement with experiment for
the bond length is obtained with the FSLYP functional, both AE and
ECP. B3LYP follows with a bond length $0.03\uAA$ longer than the one
computed with FSLYP. Becke98 bond length is in the worst agreement
with the experiment, but only $\approx 0.005\uAA$ longer than the
B3LYP bond length.

For each of the three XC functionals used, ECP calculation predicts
shorter bond length than the AE one by $\approx 0.02\uAA$, and, thus,
it is in better agreement with the experiment.

\subsubsection{Dissociation energy} \label{sssec:DE}

The computed dissociation energies span a large range of values, from
$1.33\eV$ for FSLYP/ECP to $2.07\eV$ for Becke98/AE. The deviations
from the experimental value of the computed dissociation energy,
$\Delta D_e = D_e^\text{comp}-D_e^\text{exp}$ range between
$-0.525\eV$ $(-28.4\,\%)$ and $0.221\eV$ $(11.9\,\%)$.

Among our DFT calculations, the best agreement with the experiment for
the dissociation energy of Ni$_2$ is obtained with the B3LYP
functional. B3LYP/ECP slightly overestimates $D_e$ by $0.001\eV$
$(0.1\,\%)$, while B3LYP/AE slightly underestimates $D_e$ by
$0.015\eV$ $(0.8\,\%)$. This excellent agreement with the experiment
of the B3LYP functional is clearly fortuitous since the errors in the
B3LYP dissociation energies average $0.10\eV$, with a maximum absolute
deviation of $0.36\eV$ for the G2 set of
molecules~\cite{Bauschlicher95:XC_accur_G2}.
Becke98/ECP comes second and underestimates $D_e$ by $0.058\eV$
$(3.1\,\%)$, performing only slightly worse than CASPT2, which
overestimates $D_e$ by $0.040\eV$ $(2.2\,\%)$.
FSLYP/AE is next and it underestimates $D_e$ by $0.186\eV$
$(10.1\,\%)$ similar to CASSCF/IC-ACPF, for which $\Delta D_e =
-0.159\eV$ $(-8.6\,\%)$.
Becke98/AE and FSLYP/ECP are the methods that give the largest
disagreement with the experiment: Becke98/AE overestimates  $D_e$ by
$0.221\eV$ $(11.9\,\%)$, and FSLYP/ECP underestimates $D_e$ by
$0.525\eV$ $(28.4\,\%)$.

The effects of ECP and XC functionals on the dissociation energy of
Ni$_2$ do not seem to show similar trends to the ones seen for the
bond length. However, similar trends can be noticed if, instead
of $D_e$, one compares the asymptotic dissociation energy, $D_e^a$,
which is the dissociation energy with respect to the $^3D$ atoms that
correlate with the ground state of the nickel dimer ($D_e^a = D_e +
2E_{\text{Ni }^3D}$, where $E_{\text{Ni }^3D}$ is the energy of the
$^3D$ state of Ni atom relative to the energy of the ground state).

The agreement of computed $D_e^a$ with the experimental value is
clearly better than that of $D_e$. B3LYP/AE with $\Delta D_e^a =
-0.015\eV$ $(-0.8\,\%)$ and B3LYP/ECP $\Delta D_e^a = 0.031\eV$
$(1.7\,\%)$ give the best agreement with the experiment, similar to
CASPT2, for which $\Delta D_e^a = 0.040\eV$ $(2.2\,\%)$, and FSLYP/AE
$\Delta D_e^a = 0.052\eV$ $(2.8\,\%)$.  FSLYP/ECP with $\Delta D_e^a =
0.106\eV$ $(5.7\,\%)$ is a little worse than FSLYP/AE. Becke98 gives
the largest overestimation for $D_e^a$: Becke98/AE gives $\Delta D_e^a
= 0.221\eV$ $(11.9\,\%)$ and Becke98/ECP gives $\Delta D_e^a =
0.349\eV$ $(18.8\,\%)$.

For all three functionals, the ECP basis tends to overestimate the
$D_e^a$ compared to the AE basis. For B3LYP and FSLYP the effect of
ECP on $D_e^a$ is the smallest among the three functionals ($\approx
0.05\eV$), while for the Becke98 functional the effect of ECP on
$D_e^a$ is largest ($0.13\eV$), for which $\Delta D_e^a$ increases
from $0.22\eV$ for AE to $0.35\eV$ for ECP. However, for Becke98/ECP
$\Delta D_e$ is only $-0.06\eV$ due to cancellation of large and
positive $\Delta D_e^a$ and the large $E(\text{Ni }^3D)$. For FSLYP
this cancellation doesn't happen and both FSLYP/AE and FSLYP/ECP
underestimate the dissociation energy by fairly large amount because
of the large error in $E(^3D\text{ Ni})$.

\subsubsection{Vibrational frequency} \label{sssec:VibFreq}

As can be noticed in Fig.~\ref{fig:Ni2-gs-RD}, there seem to be a
general trend for all our DFT calculations, that the error in
vibrational frequency decreases as the error in bond length
increases. CASSCF/IC-ACPF is close to following the same trend, but
CASPT2 is clearly an outlier.

It is apparent that all calculations included in
Table~\ref{tab:Ni2-gs} and Fig.~\ref{fig:Ni2-gs-RD} --- both our
DFT calculations and CASPT2 and CASSCF/IAACPF wavefunction methods
included for comparison --- overestimate the vibrational frequency of
Ni$_2$. The deviations from the experimental value of the computed
harmonic vibrational frequency, $\Delta \omega_e =
\omega_e^\text{comp} - \omega_e^\text{exp}$ range between $10.6\icm$
($4.3\,\%$) and $36.8\icm$ $(14.9\,\%)$ among our DFT results.

Becke98/AE with $\Delta \omega_e = 10.6\icm$ $(4.3\,\%)$ and B3LYP/AE
$\Delta \omega_e = 12.7\icm$ $(5.2\,\%)$ give the best agreement with
the experiment among our DFT results, slightly worse than
CASSCF/IC-ACPF, for which $\Delta \omega_e = 6.8\icm$
$(2.8\,\%)$. Becke98/ECP follows, overestimating $\omega_e$ by
$18.9\icm$ $(7.7\,\%)$. B3LYP/ECP and FSLYP/AE perform similarly with
$\Delta \omega_e = 23.1\icm$ $(9.4\,\%)$ and $\Delta \omega_e =
24.9\icm$ $(10.1\,\%)$, respectively. FSLYP/ECP with $\Delta \omega_e
= 36.8\icm$ $(14.9\,\%)$ gives the worst agreement with experiment,
similar to CASPT2, which overestimates $\omega_e$ by $34.8\icm$
$(14.1\,\%)$.

\subsubsection{Summary of the results for $d^A_\delta d^B_\delta$-holes states of Ni$_2$}

%
\begin{table}[tbp!]
\caption[DFT results for Ni$_2$.]{\label{tab:ni2-results}
DFT results for Ni$_2$.  $d_e$ -- bond length ($\uAA$), $D_e$ --
dissociation energy, relative to ground state Ni atoms (without
zero-point correction, $\eV$), $\omega_e$ -- vibrational frequency
($\icm$). The notation used for the states is $^{M}(h^Ah^B)$, where
$M$ is the multiplicity, $h^A$ and $h^B$ are the holes on Ni atoms $A$
and $B$, respectively. The $S_z=0$, $\avg{S^2}=1$ mixed states are
denoted by $^{1,3}(h^Ah^B)$.}
\begin{ruledtabular}
\begin{tabular}{llrrr}
 Method & 
 State & 
 \multicolumn{1}{c}{\mbox{$d_e$}} & 
 \multicolumn{1}{c}{\mbox{$D_e$}} & 
 \multicolumn{1}{c}{\mbox{$\omega_e$}} \\
\hline
Becke98/AE & \Tuu       & 2.302 & 2.054 & 257.0\minh \\
           & \Suv       & 2.298 & 2.068 & 256.8 \\
           & \uSuv      & 2.298 & 2.065 & 257.0 \\
           & \Tuv       & 2.297 & 2.062 & 257.1 \\
           & \Suu       & 2.296 & 2.071 & 256.8 \\
           & \uSuu      & 2.299 & 2.062 & 256.9 \\
\hline
Becke98/ECP & \Tvv      & 2.283 & 1.779 & 266.6\minh \\
           & \Tuu       & 2.282 & 1.783 & 266.6 \\
           & \Suv       & 2.280 & 1.788 & 265.3 \\
           & \uSuv      & 2.279 & 1.787 & 265.9 \\
           & \Tuv       & 2.279 & 1.787 & 266.4 \\
           & \Svv       & 2.278 & 1.787 & 265.0 \\
           & \Suu       & 2.278 & 1.792 & 265.1 \\
           & \uSvv      & 2.280 & 1.783 & 265.8 \\
           & \uSuu      & 2.280 & 1.787 & 265.9 \\
\hline
B3LYP/AE   & \Tuu       & 2.296 & 1.817 & 260.1\minh \\
           & \Suv       & 2.293 & 1.832 & 259.2 \\
           & \uSuv      & 2.292 & 1.828 & 259.6 \\
           & \Tuv       & 2.292 & 1.825 & 260.0 \\
           & \Suu       & 2.291 & 1.835 & 258.9 \\
           & \uSuu      & 2.294 & 1.826 & 259.5 \\
\hline
B3LYP/ECP  & \Tuu       & 2.275 & 1.844 & 269.4\minh \\
           & \Suv       & 2.273 & 1.848 & 267.7 \\
           & \uSuv      & 2.272 & 1.850 & 268.5 \\
           & \Tuv       & 2.271 & 1.851 & 269.3 \\
           & \Suu       & 2.271 & 1.850 & 267.6 \\
           & \uSuu      & 2.273 & 1.847 & 268.5 \\
\hline
FSLYP/AE   & \Tuu       & 2.264 & 1.645 & 272.4\minh \\
           & \Suv       & 2.262 & 1.662 & 271.3 \\
           & \uSuv      & 2.261 & 1.656 & 271.7 \\
           & \Tuv       & 2.261 & 1.650 & 272.2 \\
           & \Suu       & 2.260 & 1.664 & 271.0 \\
           & \uSuu      & 2.262 & 1.654 & 271.7 \\
\hline
FSLYP/ECP  & \Tvv       & 2.240 & 1.307 & 284.4\minh \\
           & \Tuu       & 2.240 & 1.299 & 284.4 \\
           & \Suv       & 2.238 & 1.319 & 283.3 \\
           & \uSuv      & 2.237 & 1.314 & 283.8 \\
           & \Tuv       & 2.237 & 1.309 & 284.2 \\
           & \Svv       & 2.236 & 1.325 & 283.0 \\
           & \Suu       & 2.236 & 1.318 & 283.0 \\
           & \uSvv      & 2.238 & 1.316 & 283.7 \\
           & \uSuu      & 2.238 & 1.308 & 283.7 
\end{tabular}
\end{ruledtabular}
\end{table}

All calculations predict {\mbox{$d^A_\delta d^B_\delta$-holes}} states
to have the lowest energy both for singlet and for triplet spin
multiplicities. The bond lengths of optimized geometries, dissociation
energies and vibrational frequencies for these states calculated with
the described DFT methods are tabulated in Table~\ref{tab:ni2-results}
for comparison.

The first observation is that the {\Tuu{}} and/or {\Tvv{}} are the
highest-lying states, for all calculations, and that the spin
projection changes the ordering of the singlet states for all three
all-electron calculations.  For these calculations, the lowest energy
is obtained for the unprojected singlet {\uSuv{}} state, and upon
projection, the degenerate {\Suu{}} and {\Svv{}} become the ground
state.

For the B3LYP/ECP calculation spin projection does not change the
{\Tuv{}} ground state, although it makes the {\Tuv{}} ground state
nearly degenerate with the degenerate {\Suu{}} and {\Svv{}}. However,
the {\Tuv{}} ground state is only $0.001\eV$ lower in energy than the
degenerate {\Suu{}} and {\Svv{}}.
For Becke98/ECP the unprojected ground state is {\uSuv{}} degenerate
with {\Tuv{}}, and upon spin projection, {\Suu{}} becomes the ground
state, with a dissociation energy larger than the one of {\Svv{}} by
$0.005\eV$.
For FSLYP/ECP the {\uSvv{}} unprojected ground state does not change
upon spin projection, but the difference between the $D_e$ of {\Suu{}}
and that of {\Svv{}} is the largest among all calculations:
$0.008\eV$, and is larger than the numerical accuracy of the DFT
calculations (better than $0.1\meV$).

It is also worth noting that for all calculations the average $D_e$ of
singlet states is larger than the one of the triplet states. However,
the difference between the singlet and the triplet is very small for
Becke98/ECP and B3LYP/ECP ($0.006\eV$ and $0.003\eV$,
respectively). For the other calculations, the difference is somewhat
larger, around $0.015\eV$.

However, it is important to note from Table~\ref{tab:ni2-results}
that for each combination of exchange-correlation functional and basis
set used, all {\mbox{$\delta\delta$-holes}} states are in a very
narrow energy range: $\approx 20\meV$ for all all-electron
calculations, $26\meV$ for FSLYP/ECP, $13\meV$ for Becke98/ECP and
only $7\meV$ for B3LYP/ECP.

Since, as shown above, the ordering of states can change upon
spin-projection, if possible to perform, spin-projection is desirable.
However, we want to emphasize that the differences between the lowest
broken-symmetry singlet states and the projected singlet ground
states, for the all-electron calculations and FSLYP/ECP, is less than
$10\meV$, and for some applications that difference may not be
relevant.  Nevertheless, we plan to consider spin-projection for
larger clusters, if possible, at least for evaluating the errors that
arise from it.

%
\begin{figure*}[tbp]
  \resizebox{6.75in}{!}{\includegraphics*{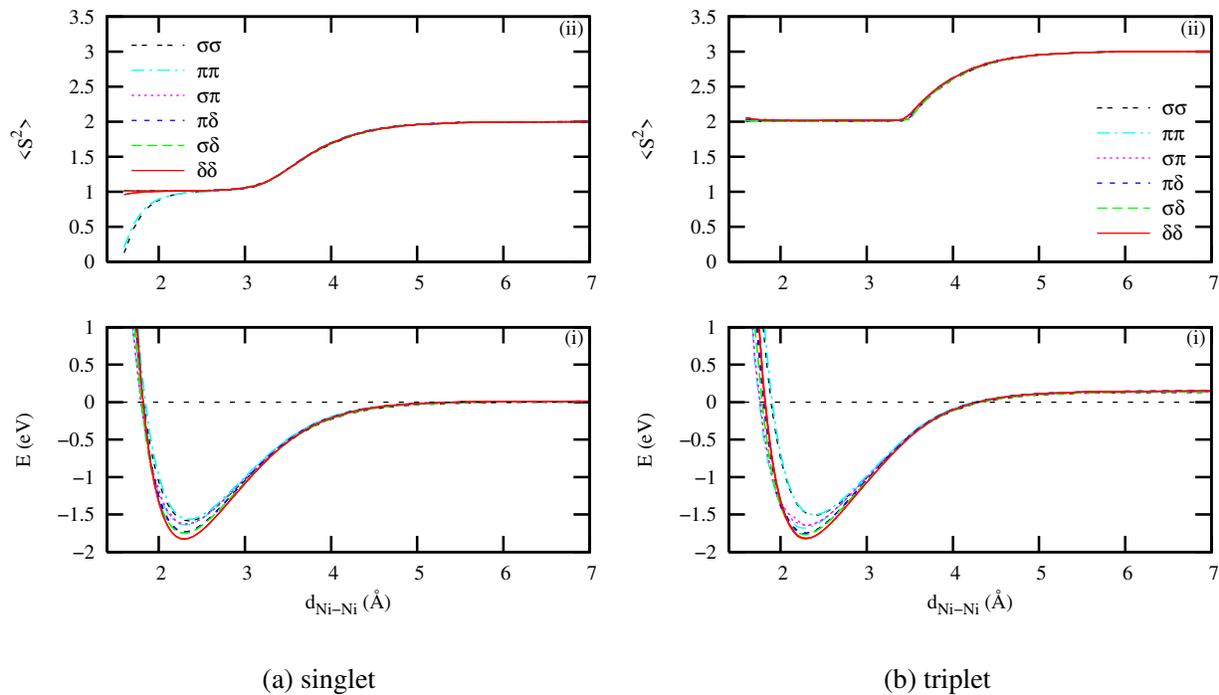}}
\caption[B3LYP/\mbox{Wachters+f} PEC's of Ni$_2$]%
{\label{fig:Ni2-pecs-b3lyp-wachtf} 
B3LYP/\mbox{Wachters+f} PEC's of Ni$_2$. Energy in eV, relative to
ground state Ni atoms and bond length in \AA{}.}
\end{figure*}

%
\begin{figure*}[tbp]
  \resizebox{6.75in}{!}{\includegraphics*{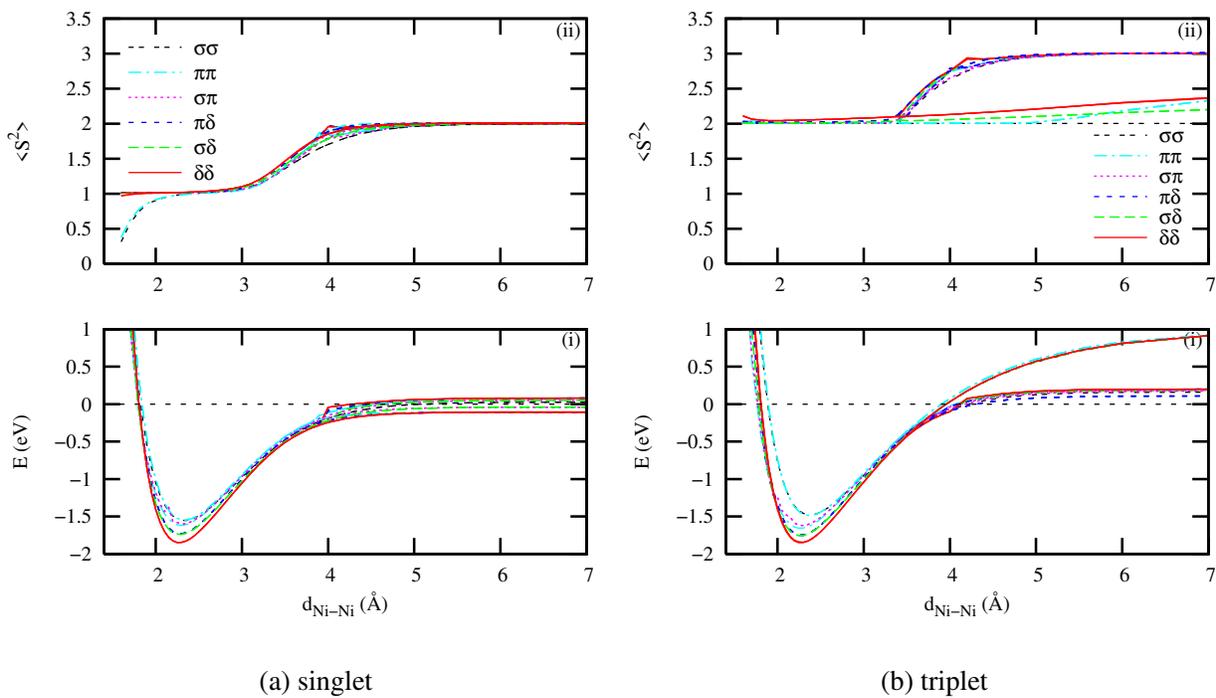}}
\caption[B3LYP/\mbox{Stuttgart RSC ECP} PEC's of Ni$_2$]%
{\label{fig:Ni2-pecs-stuttgarts} 
B3LYP/\mbox{Stuttgart RSC ECP} PEC's of Ni$_2$. Energy in eV, relative
to ground state Ni atoms and bond length in \AA{}.}
\end{figure*}

\subsubsection{Potential energy curves (PEC)}

In order to determine the ground state of Ni$_2$ we did a full scan of
the PEC for each method and for each unique combination of holes.  All
calculations predict {\mbox{$\delta\delta$-holes}} states to have the
lowest energy, with the next level $50\text{--}100\meV$ above,
$\sigma\delta$ for Becke98 and B3LYP calculations and $\pi\delta$ for
FSLYP calculations.

The computations of $\sigma\pi$ states with Becke98 and B3LYP
functionals only converge to $10^{-5}\text{--}10^{-4}\Hartree$ within
100 iterations in the $1.95\text{--}2.55\uAA$ range.  Because the
FSLYP calculations, which converge properly, predict that these states
are $\approx 200\meV$ higher, the same value as the
``not-so-converged'' results for the above calculations, we have
chosen not to investigate the matter any further.

Since the results of the PEC scans are rather similar, and B3LYP is
our functional of choice, in the following discussion of the PEC's, we
focus attention principally on the results from B3LYP calculations.

The B3LYP/AE and B3LYP/ECP potential energy curves (PEC) of singlet
($S_z=0$) and triplet ($S_z=1$) states of Ni$_2$ (both unrestricted,
symmetry broken) are shown in Fig.~\ref{fig:Ni2-pecs-b3lyp-wachtf} and
in Fig.~\ref{fig:Ni2-pecs-stuttgarts}, respectively, along with the
variation of $\avg{S^2}$ with the bond length for all possible
positions of holes in the $3d$ shell on both atoms, grouped by hole
type.  The first trend that can be noticed is that the equilibrium
bond length increases as the dissociation energy decreases.  Aside for
a few states (singlet $\sigma\sigma$ and $\pi\pi$), all states have
$\avg{S^2}\approx 1$ over a large interval, validating the weakly
interacting $3d^9$ units model for a large range of bond lengths.
Even the singlet $\sigma\sigma$ and $\pi\pi$ states have
$\avg{S^2}\approx 1$ in a range of about $\pm 1\uAA$ around the
equilibrium bond length.

One can notice a big difference between AE and ECP PEC's: ECP PEC's
branch around $3.5\uAA$.  There are two causes for branching: one,
which is not related to functional~\footnote{For the exact
Hohenberg-Kohn (HK) functional there would be no unrestricted
solution, but all of the current approximations of the HK functional
suffer from this problem~\cite{Cremer:01}.} or ECP, is the
restricted--unrestricted crossover, while the other cause is
dissociation into $^3F$ ground state of Ni atoms.  These two effects
overlap because the branching is obtained by scanning the PEC from
$\approx{}3.5\uAA$, increasing the bond length and using as initial
guess the molecular orbitals from the previous calculation.  Depending
on the initial guess, the calculation may end in the restricted or
unrestricted solution, or, at large distances, the calculation may
converge to the $^3F+^3F$, $^3F+^3D$ or $^3D+^3D$ states of the Ni
atoms.  The restricted-unrestricted branching is likely to show up for
any of the methods, but the ground-state branching can only appear for
the methods that predict $^3F$ ground state for Ni, namely FSLYP/AE
FSLYP/ECP, and Becke98/ECP along with the discussed B3LYP/ECP.

One can also notice that some of the B3LYP/ECP PEC's have asymptotes
below 0, \ie, below the energy of the ground state of the two
nickel atoms. A closer look reveals that the asymptotes of
the $\dxy\dxz$-, $\dxy\dyz$-, $\dxz\dyz$-, $\dxz\dxz$-, $\dyz\dyz$-,
$\dxy\dxy$-, $\dxsqmysq\dxz$-, $\dxsqmysq\dyz$-, $\dxsqmysq\dxy$-, and
$\dxsqmysq\dxsqmysq$-holes states lie $0.1\eV$ below the ground state
of the two nickel atoms, the ones of $\dzsq\dxz$-, $\dzsq\dyz$-,
$\dzsq\dxy$-, and $\dzsq\dxsqmysq$-holes states lie $0.04\eV$ below the
ground state of the two nickel atoms, and only $\dzsq\dzsq$-holes
state lies $0.03\eV$ above the ground state of the two nickel
atoms. The most likely explanation for this observation is that
B3LYP/ECP predicts a lower energy for a state that is not in the space
of states spanned by our initial guess. This issue needs further
investigation, but since the effect is rather small (at most
$0.05\eV$/nickel atom), we chose to investigate the issue in a further
paper.

The initial PEC scans are done either with broken symmetry atomic
initial guess ($3d^94s^1\uparrow\uparrow+\downarrow\downarrow3d^94s^1$
for singlet and $3d^94s^1\uparrow\uparrow+\downarrow\uparrow3d^94s^1$
for triplet) at each bond length or, starting from $10\uAA$ and
decreasing the bond length and using as initial guess the molecular
orbitals at the previous bond length.  Either initial guess gives the
same results, but the method using atomic initial guess needs a few
extra iterations.  For larger cluster calculations it may be useful to
save the molecular orbitals at each geometry configuration and try to
reuse them for a neighboring point calculation.

It is apparent from Fig.~\ref{fig:Ni2-pecs-b3lyp-wachtf} that for the
B3LYP/AE calculation the singlet dissociates to the correct $2{}^3D$
atoms limit ($\avg{S^2}=2$), whereas the triplet dissociates to
${}^3D+{}^{1,3}D$, which is $0.14\eV$ above the correct limit.  This
type of error only plays an important role at large distances, when
the molecule starts to resemble two separated atoms, and can be
correlated with $\avg{S^2}$ of the Kohn-Sham determinant.  When
$\avg{S^2}$ is close to the exact value, this type of error is not
present.  For Ni$_2$, both singlet and triplet, the $\avg{S^2}$ is
correct (i.e. equal to the theoretical value) for $2\uAA<d_e<3\uAA$.
At interatomic distances greater than approximately $3\uAA$,
$\avg{S^2}$ starts to increase, and so does the error in the energy of
the triplet.  At interatomic distances larger than approximately
$\approx 4\uAA$, $\avg{S^2}$ for the triplet reaches a value of
$\approx 3$ and stays constant for larger distances.  Similarly, the
error in the energy of the triplet approaches the asymptotic value of
$0.14\eV$.  

In larger clusters, this could be a potential issue for computing
barriers.  However, only configurations in which one atom is at
sufficiently large distance from other atoms, completely or partly
detached (evaporated) from the cluster, and in the ${}^{1,3}D$ state,
would encounter the above described problem.  Moreover, the error
(\mbox{$\le{}0.14\eV$}) could be important if the height of the
barrier were small.  But the evaporation energy of an atom from the
cluster is likely to be of the same order of magnitude as the
dissociation energy of the dimer (\mbox{$\approx{}1.5\eV$}), and 
the height of the barrier would be overestimated by
\mbox{$\approx{}10\,\%$}.  Consequently, this error should be
unimportant for large clusters.


\subsection{Nickel Hydride} \label{ssec:hydride}

%
\begin{table*}[tbp]
\begin{minipage}{\textwidth}
\caption[Ground state of NiH -- comparison between computations and
experiment.]{\label{tab:NiH-gs}
Ground state of NiH -- comparison between computations and experiment.
$d_e$ -- bond length ($\uAA$), $D_e$ -- dissociation energy, relative to
ground state Ni atoms (without zero-point correction, $\eV$),
$\omega_e$ -- vibrational frequency ($\mathrm{cm}^{-1}$) and $\mu$ --
dipole moment (Debye). The relative errors with respect to
experimental values are given in parentheses, and the average (AARD)
and the maximum (MARD) absolute relative deviations from experimental
values of $d_e$, $D_e$, $\omega_e$ and $\mu$ are listed under the AARD
and MARD columns, respectively.}
\begin{ruledtabular}
\begin{tabular}{l@{\hspace*{1.5em}}l@{}r@{\hspace*{1.5em}}l@{}r@{\hspace*{1.5em}}l@{}r@{\hspace*{1.5em}}l@{}r@{\hspace*{1.5em}}d@{\hspace*{1.5em}}d}
 Method
 & \multicolumn{2}{c@{\hspace*{1.5em}}}{\mbox{$d_e$}}
 & \multicolumn{2}{c@{\hspace*{1.5em}}}{\mbox{$D_e$}}
 & \multicolumn{2}{c@{\hspace*{1.5em}}}{\mbox{$\omega_e$}}
 & \multicolumn{2}{c@{\hspace*{1.5em}}}{\mbox{$\mu$}}
 & \multicolumn{1}{c@{\hspace*{1.5em}}}{\mbox{AARD}}
 & \multicolumn{1}{c}{\mbox{MARD}}
\\
\hline
B3LYP/ECP   &  1.454 & (-1.6) & 2.901 & (13.8) & 1937.6 & (-0.2) & 2.29 & (-12.7) &  7.1 & 13.8\minh \\
Becke98/ECP &  1.456 & (-1.4) & 2.808 & (10.1) & 1927.6 & (-0.7) & 2.43 &  (-7.2) &  4.8 & 10.1 \\
B3LYP/AE    &  1.474 & (-0.2) & 2.856 & (12.0) & 1940.2 & (-0.1) & 2.43 &  (-7.1) &  4.8 & 12.0 \\
FSLYP/AE    &  1.470 & (-0.5) & 2.681 &  (5.1) & 1943.8 &  (0.1) & 2.91 &  (11.2) &  4.2 & 11.2 \\
CASPT2\footnote{We report here the values from Table~VI of
Ref.~\mbox{\onlinecite{Pou-Amerigo:94}} for $d_e$, $D_e$ and
$\omega_e$, and from Table~VII for $\mu$ [PT2F($3s3p$)+RC], from which
we subtract the estimated relativistic corrections (RC). From the same
reference, we estimate the RC to $d_e$ and $D_e$ from Table~V and the
RC to $\mu$ from Table~VII, as the difference between the PT2F+RC
values and the PT2F ones. We use the MRCI RC to $\omega_e$ from
Ref.~\mbox{\onlinecite{Marian:89:NiH-MR-RC}}. We also subtract these
relativistic corrections from the experimental values.}
            &  1.463 & (-0.9) & 2.76  &  (8.2) & 2022.3 &  (4.2) & 2.54 &  (-3.1) &  4.1 &  8.2 \\
Becke98/AE  &  1.477 &  (0.0) & 2.888 & (13.3) & 1944.2 &  (0.1) & 2.59 &  (-1.0) &  3.6 & 13.3 \\
FSLYP/ECP   &  1.449 & (-1.9) & 2.526 & (-0.9) & 1953.4 &  (0.6) & 2.74 &   (4.5) &  2.0 &  4.5 \\
Exp.\footnote{Experimental values from which we subtract the CASPT2
relativistic corrections (RC) to $d_e$, $D_e$ and $\mu$ from
Ref.~\mbox{\onlinecite{Pou-Amerigo:94}}, and the MRCI RC to $\omega_e$
from Ref.~\mbox{\onlinecite{Marian:89:NiH-MR-RC}} (see
footnote~\NiHRC). The experimental value of $d_e$ is
$1.454\uAA$~\citep[cited in
Ref.~\mbox{\onlinecite{Pou-Amerigo:94}}]{Gray:91:NiH}, from which we
subtract the CASPT2 RC of $-0.023\uAA$; the experimental value of $D_e$
is $2.70\eV$~\citep[recommended value from Ref.][cited in
Ref.~\mbox{\onlinecite{Pou-Amerigo:94}}]{Armentrout:92:TM_hydrid},
from which we subtract the CASPT2 RC of $0.15\eV$; the experimental
value of $\omega_e$ is $2001.3\unit{cm}^{-1}$~\citep[cited in
Ref.~\mbox{\onlinecite{Pou-Amerigo:94}}]{Gray:91:NiH}, from which we
subtract the MRCI RC of $60\unit{cm}^{-1}$; the experimental value of
$\mu$ is $2.4\pm{}0.1\unit{Debye}$~\citep[cited in
Ref.~\mbox{\onlinecite{Pou-Amerigo:94}}]{Gray:85:NiH-mu}, from which
we subtract the CASPT2 RC of $-0.22\unit{Debye}$.}
            &  1.477 &        & 2.55  &        & 1941.3 &        & 2.6   &   (3.8) &      &      \\
\end{tabular}
\end{ruledtabular}
\end{minipage}
\end{table*}

%
\begin{figure}[tbp]
	\resizebox{3.375in}{!}{\includegraphics*{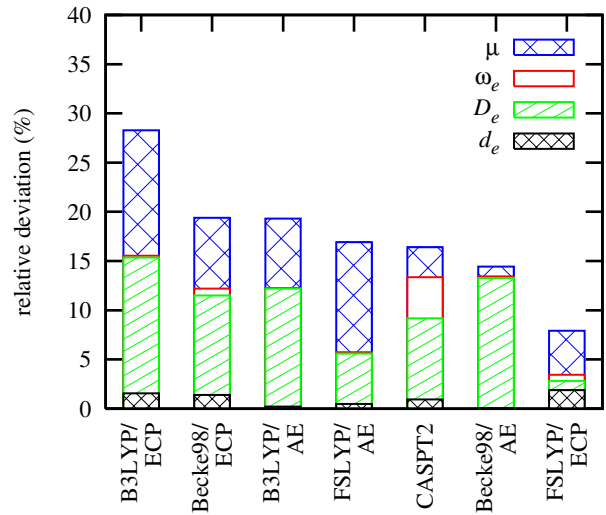}}%
\caption[NiH: performance of the used methods]%
{\label{fig:NiH-gs-ARD} The absolute relative deviations from
experiment of computed dissociation energy ($D_e$), bond length
($d_e$), vibrational frequency ($\omega_e$), and dipole moment ($\mu$)
for NiH. The results are arranged from left to right in order of
decreasing total absolute relative deviation (TARD) --- the sum of
absolute relative deviations from the experimental values of the
computed $d_e$, $D_e$, $\omega_e$ and $\mu$.}
\end{figure}

The bond lengths ($d_e$), dissociation energies ($D_e$),
vibrational frequencies ($\omega_e$) and dipole moment ($\mu$) 
for ground states of NiH from different calculations are reported in
Table~\ref{tab:NiH-gs} along with experimental values and results from
other theoretical studies, listed in the order of decreasing average
of absolute relative deviations (AARD) from the experimental values of
the computed $d_e$, $D_e$, $\omega_e$ and $\mu$.

The experimental values reported in Table~\ref{tab:NiH-gs} are the
deperturbed values of $d_e$ and $\omega_e$ of Gray \etal~\citep[cited
in Ref.~\mbox{\onlinecite{Pou-Amerigo:94}}]{Gray:91:NiH}, the
recommended value of $D_e$ from
Ref.~\onlinecite{Armentrout:92:TM_hydrid} (cited in
Ref.~{\onlinecite{Pou-Amerigo:94}}) and $\mu$ from
Ref.~\onlinecite{Gray:85:NiH-mu} (cited in
Ref.~{\onlinecite{Pou-Amerigo:94}}), from which we subtract the CASPT2
relativistic corrections (RC) to $d_e$, $D_e$ and $\mu$ from
Ref.~\mbox{\onlinecite{Pou-Amerigo:94}}, and the MRCI RC to $\omega_e$
from Ref.~\mbox{\onlinecite{Marian:89:NiH-MR-RC}} (see
footnote~\NiHRC\ of Table~\ref{tab:NiH-gs} for details).

The absolute relative deviations from the experimental values of the
computed bond lengths ($d_e$), dissociation energies ($D_e$),
vibrational frequencies ($\omega_e$) and dipole moment ($\mu$) of the
ground state of NiH are plotted in Fig.~\ref{fig:NiH-gs-ARD} for
comparison. They are arranged from left to right in order of
decreasing total absolute relative deviation (TARD) --- the sum of
absolute relative deviations from the experimental values of the
computed $d_e$, $D_e$, $\omega_e$ and $\mu$.

From Fig.~\ref{fig:NiH-gs-ARD}, as well as from
Table~\ref{tab:NiH-gs}, it is apparent that for NiH, the best overall
agreement with experiment among our DFT calculations is obtained for
FSLYP/ECP ($7.9\,\%$ TARD), followed by Becke98/AE ($14.4\,\%$ TARD)
and FSLYP/AE ($16.9\,\%$ TARD) similar to CASPT2 ($16.4\,\%$
TARD). B3LYP/AE ($19.3\,\%$ TARD) is next, similar to Becke98/ECP
($19.4\,\%$ TARD), and B3LYP/ECP ($28.3\,\%$ TARD) gives the largest
disagreement with experiment.

All our DFT calculations predict $^2\Delta$ (\mbox{$\delta$-hole})
ground state, in agreement with the CASPT2 calculation and
experiment. However, it is important to note, that, for Becke98/ECP
and FSLYP/ECP results the difference between the $\dxy$-hole and the
$\dxsqmysq$-hole components of the $^2\Delta$ state --- $2\meV$ and
$5\meV$, respectively --- is larger than the error of the DFT
calculations ($\leq 0.1\meV$). We report the energy of the component
with the lowest energy as the energy of the ground state.

%
\begin{figure}[tbp]
	\resizebox{3.375in}{!}{\includegraphics*{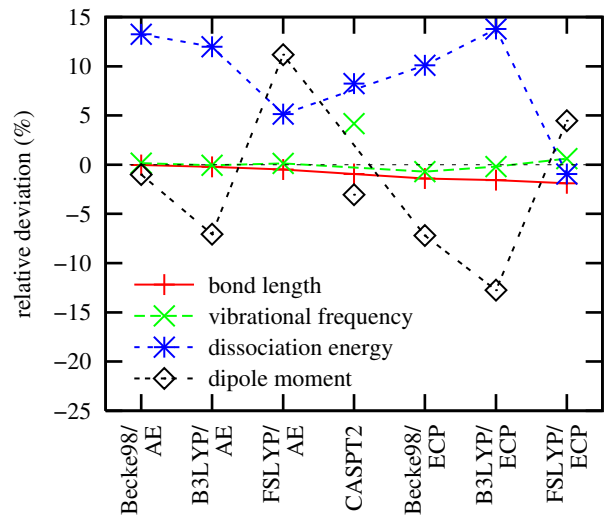}}%
\caption[Signed errors for NiH]%
{\label{fig:NiH-gs-RD} The relative deviations of computed bond
length, dissociation energy, vibrational frequency and dipole moment
from experimental values for NiH. Only the results from our DFT
calculations are connected by lines. CASPT2 values are included for
comparison.}
\end{figure}

It is apparent that all calculations included in
Table~\ref{tab:NiH-gs} and Fig.~\ref{fig:NiH-gs-RD}
underestimate the bond length of NiH.  Becke98/AE with $\Delta d_e =
d_e^\text{comp} - d_e^\text{exp} = -0.0003\uAA$ $(-0.02\,\%)$ gives
the best agreement with experiment. The other DFT calculations and
CASPT2 give significantly shorter bond lengths for NiH than
Becke98/AE, but they can still be considered in good agreement with
the experiment, giving $\Delta d_e$ ranging from $-0.003\uAA$
$(-0.2\,\%)$ for B3LYP/AE to $-0.028\uAA$ $(-1.9\,\%)$ for FSLYP/ECP.
Among all three XC functionals, the best agreement with experiment for
the bond length is obtained with the Becke98 functional, both AE and
ECP. B3LYP follows with a bond length $0.003\uAA$ shorter than the one
computed with Becke98. FSLYP bond length is in the worst agreement
with the experiment, but only $\approx 0.004\uAA$ longer than the
B3LYP bond length.
For each of the three XC functionals used, ECP calculation predicts
shorter bond length than the AE one by $\approx 0.02\uAA$, like in the
case of Ni$_2$, but this worsens the agreement with the experiment,
unlike in the case of Ni$_2$.

The computed dissociation energies span a large range of values, from
$2.53\eV$ for FSLYP/ECP to $2.90\eV$ for B3LYP/ECP. Among all DFT
computations, only FSLYP/ECP underestimates $D_e$ by $0.024\eV$
$(0.9\,\%)$, and gives the best agreement with the experiment.  All
other DFT computations and CASPT2 overestimate $D_e$: FSLYP/AE by
$5\,\%$, CASPT2 by $8\,\%$, and Becke98/ECP, B3LYP/AE, Becke98/AE, and
B3LYP/ECP by $10\,\%$, $12\,\%$, $13\,\%$, and $14\,\%$, respectively,
giving the largest disagreement with experiment.
Like in the case of Ni$_2$, he effects of ECP and XC functionals on
the dissociation energy of Ni$_2$ do not seem to show similar trends
to the ones seen for the bond length. It can be verified that trends
show up upon correcting $D_e$ with the energy of the $^3D$ state of
Ni, but, since for NiH there is no physical ground for that kind of
correction, we chose not to do it. However, it is worth noting that
the errors in the atomic energies have such large influence on the
energetics of molecules.

The differences in the theoretical harmonic vibrational frequencies
compared to the experimental values are less than $1\,\%$ for our DFT
calculations, while CASPT2 has the largest difference from the
experimental value among the results plotted in
Fig.~\ref{fig:NiH-gs-RD} and listed in Table~\ref{tab:NiH-gs}.

For NiH the dipole moment can be expected to be a more sensitive
measure of the quality of the method~\cite{Pou-Amerigo:94}, and a
comparison of the theoretical and experimental values of the dipole
moment listed in Table~\ref{tab:NiH-gs} and plotted in
Fig.~\ref{fig:NiH-gs-RD} shows that Becke98/AE gives the best
agreement, similar to CASPT2. B3LYP/AE underestimates the dipole
moment by $7\,\%$ and FSLYP/AE overestimates it by a large amount
($11\,\%$). ECP have a strong effect on $\mu$, lowering its value by
$\approx 0.15\,D$ ($6\,\%$), bringing FSLYP/ECP in closer agreement
with experiment and worsening the agreement for B3LYP and Becke98. It
is worth noting that Becke98 predicts a value for $\mu$ in better
agreement with the experiment than B3LYP. Since $\mu$ is a
one-electron property, this may be an indication that Becke98 gives a
more accurate ground state electron density.

\section{Conclusion} \label{sec:conclusions}

We have used DFT with hybrid exchange-correlation functionals in the
broken-symmetry unrestricted formalism to study the electronic
structure of nickel dimer and nickel hydride as model systems for
larger bare/hydrogenated nickel clusters.
We have examined three hybrid functionals: the popular B3LYP, Becke's
newest optimized functional Becke98, and the simple FSLYP functional
($50\,\%$ Hartree-Fock and $50\,\%$ Slater exchange and LYP
gradient-corrected correlation functional) with two basis sets:
all-electron (AE) Wachters+f basis set and Stuttgart RSC effective
core potential (ECP) and basis set.

For Ni$_2$, all of our DFT calculations give bond lengths that are
within $0.1\uAA$ ($5\,\%$) from the experimental value, and in good
agreement with the high-level wavefunction methods
CASPT2\cite{Pou-Amerigo:94} and CASSCF/IC-ACPF\cite{Bauschlicher:92}.
Only Becke98/AE and B3LYP/AE give harmonic vibrational frequencies
that are within $5\,\%$ from the experimental value, similar to
CASSCF/IC-ACPF. Becke98/ECP, B3LYP/ECP and FSLYP/AE give $\omega_e$
within $10\,\%$ from the experimental value, similar to CASPT2, and
FSLYP/ECP overestimates the experimental $\omega_e$ by $15\,\%$.  The
discrepancies between calculated and experimental values of
dissociation energy span a large range, between $-28\,\%$ and
$12\,\%$. B3LYP/ECP, B3LYP/AE and Becke98/ECP give values of $D_e$
that are within less than $5\,\%$ from the experimental value, similar
to CASPT2. FSLYP/AE and Becke98/AE give values of $D_e$ that are a
within $12\,\%$ from experimental value, similar to
CASSCF/IC-ACPF. FSLYP/ECP gives a value of $D_e$ that is smaller than
the experimental value by $28\,\%$.

For NiH, all of our DFT calculations give bond lengths that are within
$0.03\uAA$ ($2\,\%$) from the experimental value, and in good
agreement with CASPT2\cite{Pou-Amerigo:94}. They also give harmonic
vibrational frequencies that are within less than $15\icm$ ($1\,\%$)
from the experimental value, in better agreement with experiment than
CASPT2, which overestimates $\omega_e$ by $4\,\%$. The discrepancies
between the calculated and the experimental values of dissociation
energy span a large range for NiH like they do for Ni$_2$. FSLYP/ECP
underestimates $D_e$ by $1\,\%$, giving the best agreement with the
experiment. All other DFT calculations and CASPT2 overestimate $D_e$
by amounts between $5\,\%$ and $15\,\%$. For the dipole moment the
deviations from the experimental value span the largest range: between
$-13\,\%$ for B3LYP/ECP and $11\,\%$ for FSLYP/AE. Underestimating it
by $1\,\%$, Becke98/AE gives the best agreement with the experiment
for the dipole moment of NiH, similar to CASPT2, which underestimates
it by $3\,\%$.

We also find that for Ni$_2$, the spin-projection for the
broken-symmetry unrestricted singlet states changes the ordering of
the states, but the splittings are less than $10\meV$.  All our
calculations predict a {\mbox{$\delta\delta$-hole}} ground state for
Ni$_2$ and {\mbox{$\delta$-hole}} ground state for NiH.  Upon
spin-projection of the singlet state of Ni$_2$, almost all of our
calculations: Becke98 and FSLYP both AE and ECP and B3LYP/AE predict
{\Suu{}} or {\Svv{}} ground state, which is a mixture of {\SSgp{}} and
{\SGg{}}. B3LYP/ECP predicts a {\Tuv{}} (mixture of {\TSgm{}} and
{\TGu{}}) ground state virtually degenerate with the {\Suu{}}/{\Svv{}}
state, which is only $1\meV$ higher in energy than {\Tuv{}} ground
state. The doublet {\mbox{$\delta$-hole}} ground state of NiH
predicted by all our calculations is in agreement with the
experimentally predicted $^2\Delta$ ground state.  For Ni$_2$, all our
results are consistent with the experimentally predicted ground state
of $0_g^+$ (a mixture of $\SSgp{}$ and $\TSgm{}$) or $0_u^-$ (a
mixture of $\SSum{}$ and $\TSup{}$).

The goal of this paper is to establish what might comprise a minimally
reliable method for more extensive nickel cluster calculations.
Since none of the studied methods gives a good agreement with
experiment for all computed molecular properties of Ni$_2$ and NiH, we
devise an \adhoc\ quality indicator that we name overall discrepancy,
$Q$, and we calculate it with the formula:
\begin{equation}\label{eq:Q}
   Q = \underbrace{\frac17\sum_i|\epsilon_i|}_{Q_A}
      +\underbrace{\frac1{21}\sum_{i<j}\left|\epsilon_i-\epsilon_j\right|}_{Q_D},
\end{equation}
where $i$ runs over all 7 computed molecular properties for Ni$_2$ and
NiH ($d_e$, $D_e$ and $\omega_e$ of both Ni$_2$ and NiH, and $\mu$ of
NiH); $\epsilon_i$ is the relative deviation from the experimental
value of the molecular property $i$; $i<j$ stands for $i,j$ running
over all 21 unique pairs.

%
\begin{figure}[tbp]
	\resizebox{3.375in}{!}{\includegraphics*{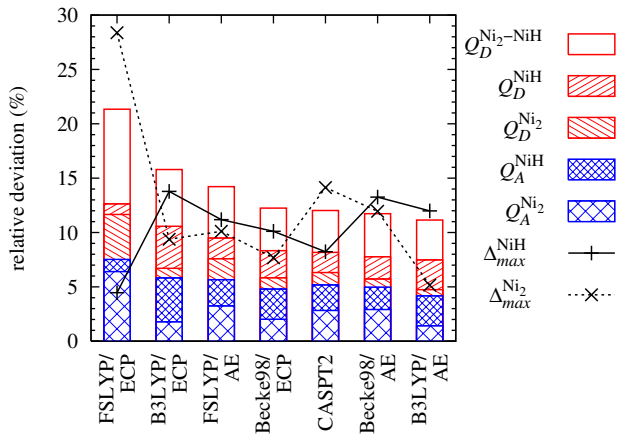}}%
\caption[Overall performance of the studied DFT
methods.]{\label{fig:tot-errs} Overall performance of the studied DFT
methods. Total values of the overall discrepancy Q are
plotted, as the total heights of the bars, along with its components
(see text for definition). Maximum absolute relative deviations from
experimental values for all computed molecular properties of Ni$_2$
($\Delta_{max}^{\text{Ni}_2}$) and NiH ($\Delta_{max}^{\text{NiH}}$)
are also shown.}
\end{figure}

%
\begin{table*}[tbp]
\caption[Charge density fit errors for ECP.]{\label{tab:ecp-cdfit-errs}
Charge density (CD) fitting errors for the bond lengths ($d_e$),
dissociation energies ($D_e$), and harmonic vibrational frequencies
($\omega_e$) of Ni$_2$ and NiH, and dipole moment ($\mu$) computed
with B3LYP and FSLYP functionals using ``Stuttgart RSC ECP'' ECP and
basis set with ``Ahlrichs Coulomb Fitting'' basis. The results from
the calculations using CD fitting are reported in the ``cdfit''
columns, the result from the calculations not using CD fitting are
reported in ``nocdfit'' columns, and the differences between the
results from the calculations using CD fitting and the results from
the ones not using CD fitting are reported under the ``cdfit err''
columns, with the percent relative errors in parentheses.}
\begin{ruledtabular}
\begin{tabular}{llrrrr|rrrr|rrrr|rrrr}
 &
 & \multicolumn{4}{c|}{\mbox{$d_e (\uAA)$}}
 & \multicolumn{4}{c|}{\mbox{$D_e (\eV)$}}
 & \multicolumn{4}{c|}{\mbox{$\omega_e (\icm)$}}
 & \multicolumn{4}{c}{\mbox{$\mu (\unit{Debye})$}}
\\
   \cline{3-6} \cline{7-10} \cline{11-14} \cline{15-18}
   Mol
 & XC
 & \multicolumn{1}{c}{\mbox{nocdfit}}
 & \multicolumn{1}{c}{\mbox{cdfit}}
 & \multicolumn{2}{c|}{\mbox{cdfit err}}
 & \multicolumn{1}{c}{\mbox{nocdfit}}
 & \multicolumn{1}{c}{\mbox{cdfit}}
 & \multicolumn{2}{c|}{\mbox{cdfit err}}
 & \multicolumn{1}{c}{\mbox{nocdfit}}
 & \multicolumn{1}{c}{\mbox{cdfit}}
 & \multicolumn{2}{c|}{\mbox{cdfit err}}
 & \multicolumn{1}{c}{\mbox{nocdfit}}
 & \multicolumn{1}{c}{\mbox{cdfit}}
 & \multicolumn{2}{c}{\mbox{cdfit err}}\minh
\\
\hline
NiH    & B3LYP &   1.454 & 1.456 & 0.002 &   (0.1) &   2.901 & 2.784 & -0.117 &  (-4.0) &  1937.6 & 1998.8 &  61.2 & (3.2)   & 2.29    & 2.21  & -0.08 & (-3.5)\minh \\
NiH    & FSLYP &   1.449 & 1.451 & 0.002 &   (0.1) &   2.526 & 2.406 & -0.120 &  (-4.8) &  1953.4 & 2015.0 &  61.6 & (3.2)   & 2.74    & 2.65  & -0.09 & (-3.3)  \\
Ni$_2$ & B3LYP &   2.271 & 2.278 & 0.007 &   (0.3) &   1.851 & 1.557 & -0.294 & (-15.9) &   269.3 &  255.3 & -14.0 & (-5.2)  &         &       &       &         \\
Ni$_2$ & FSLYP &   2.236 & 2.241 & 0.005 &   (0.2) &   1.325 & 0.999 & -0.326 & (-24.6) &   283.0 &  268.4 & -14.6 & (-5.2)  &         &       &       &         \\
\end{tabular}
\end{ruledtabular}
\end{table*}

The overall discrepancy $Q$ is the sum of two contributions: the
average discrepancy $Q_A$, which measures the overall (average)
deviation of the computed molecular properties from the experimental
values, and the consistency $Q_D$, which measures the consistency of
the methods both when computing different molecular properties of the
same molecule (\eg, $d_e$ and $\omega_e$ of Ni$_2$), and when
computing molecular properties for different molecules (\eg, $d_e$ of
Ni$_2$ and $d_e$ of NiH). For analysis, we calculate each of the
indicators $Q$, $Q_A$ and $Q_D$ for each of the molecules, by
partitioning Eq.~\ref{eq:Q} into the components for Ni$_2$
($Q^{\text{Ni}_2}$, $Q_A^{\text{Ni}_2}$, and $Q_D^{\text{Ni}_2}$), the
components for NiH ($Q^{\text{NiH}}$, $Q_A^{\text{NiH}}$, and
$Q_D^{\text{NiH}}$), and the mixed components of $Q_D$,
$Q_D^{\text{Ni}_2-\text{NiH}} =
\frac1{21}\sum_{i<j}\left|\epsilon_i-\epsilon_j\right|$ with $i$ 
running over the molecular properties of Ni$_2$ and $j$ running over
the ones of NiH.

In Fig.~\ref{fig:tot-errs} we plot the overall discrepancy $Q$ along
with its components, and the maximum absolute relative deviations from
experimental values (MARD) for all computed molecular properties of Ni$_2$
($\Delta_{max}^{\text{Ni}_2}$) and NiH ($\Delta_{max}^{\text{NiH}}$).

Fig.~\ref{fig:tot-errs} reveals that B3LYP/AE gives the lowest overall
discrepancy ($Q = 11.2\,\%$), but followed closely by Becke98/AE and
Becke98/ECP with a value of $Q$ larger than the one of B3LYP/AE by
only $0.5\,\%$ and $1\,\%$, respectively. They are also at the same
overal quality as CASPT2, for which $Q=12.0\,\%$. FSLYP/AE, with
$Q=14.2\,\%$ is a little worse than B3LYP/AE and Becke98/AE. It is
apparent from Fig.~\ref{fig:tot-errs} that the use of ECP worsen the
overall agreement with experiment. The largest effect of the ECP's is
on the results obtained with the FSLYP functional, increasing the
value of $Q$ by $7.1\,\%$. The effect is much smaller on B3LYP,
increasing $Q$ by $4.6\,\%$, and negligible on Becke98 ($0.5\,\%$).

It can be noticed that for most of the calculations included in
Fig.~\ref{fig:tot-errs}, the value of $Q$ is close to the values of
MARD for both NiH and Ni$_2$. Two methods for which that is not the
case are worth mentioning: B3LYP/AE and FSLYP/ECP. Both perform
significantly better for one of the molecules than for the other,
probably by accident. B3LYP/AE performs clearly better for Ni$_2$ than
for NiH, but its MARD for NiH agrees with $Q$, while FSLYP/ECP
performs much better for NiH than for Ni$_2$, and its MARD for Ni$_2$
is significantly larger than $Q$ (by $7.1\,\%$). Thus, FSLYP/ECP is
the only method that is not advisable to use for bare/hydrogenated
nickel clusters. However, we want to emphasize that the methods that
give the best agreement with experiment and CASPT2, B3LYP/AE,
Becke98/AE and Becke98/ECP are the methods of choice.

Our results indicate that DFT, with the B3LYP (using the Wachters+f
all-electron basis set) and Becke98 (using either Wachters+f
all-electron basis set or Stuttgart RSC effective core potential and
basis set) hybrid exchange-correlation functionals in the
broken-symmetry unrestricted formalism, becomes both an efficient and
reliable method for predicting electronic structure of our model
Ni$_2$ and NiH systems, although it is far from being a black box
method.

\begin{acknowledgments}
The authors gratefully acknowledge support from the National Science
Foundation through equipment award CHE-01-31114 and from DOE grant
DE-FG02-03ER46074.

Electronic structure calculations were performed with NWChem Versions
4.0.1, 4.1 and 4.5 (Version 4.1 with Becke98 functional patched), as
developed and distributed by Pacific Northwest National Laboratory,
P.O. Box 999, Richland, WA 99352, and funded by the U.S. Department of
Energy.

The authors gratefully acknowledge the support of Brown University's
Center for Advanced Scientific Computation and Visualization in the
present research.

One of the authors (C. V. D.) would also like to thank Dr. Richard
L. Martin for useful suggestions, and Dr. Cristian Predescu and
Dr. Dubravko Sabo for helpful discussions.
\end{acknowledgments}

\appendix

\section{Accuracy of charge density fitting}\label{sec:Accur:CDfit}

As stated in Section~\ref{sec:methods} we use charge density fitting
for the calculations using the all-electron Wachters+f basis set,
for which we employ the Ahlrichs Coulomb Fitting~\cite{Ahlrichs_cd:95,
Ahlrichs_cd:97} basis set.  For evaluating the error introduced by
charge density fitting we perform the atomic computations with B3LYP
functional and Wachters+f basis set with and without charge density
fitting.  The charge density fitting lowers the total energies of
computed atomic states by $2.5\mbox{--}3\cdot 10^{-4}\Hartree$.  The
errors in the relative energies are less severe, ranging from $-3.8$
to $1.5\meV$.

%
\begin{table}[tbp]
\caption[Charge density fit errors for Ni$_2$ (B3LYP/AE).]{\label{tab:cdfit-errs} 
Averages of charge density fit errors for B3LYP/AE optimizations and
frequency calculations for the 12 $(d_{\delta}^Ad_{\delta}^B)$ singlet
and triplet states of Ni$_2$ computed with and without charge density
fitting.  $d_e$ (m\AA) -- bond length, $E_e$ (mHartree) -- total
energy, $\omega_e$ ($\icm$) -- vibrational frequency, $E_e$ (meV) --
relative energies with respect to the ground state Ni atom,
$\Delta{}E_e$ (meV) -- relative energies with respect to the lowest
energy state from each type of calculation.  Mean -- mean of the
differences between the computations with charge density fitting and
those without; Stdev -- standard deviation of the differences; Max --
maximum absolute difference and RMS -- the root-mean-square of the
differences.}
\begin{ruledtabular}
\begin{tabular}{lccccc}
      & $d_e$&   $E_e$  & $\omega_e$& $E_e$ & $\Delta E_e$ \\
      & m\AA & mHartree & $\icm$ &   meV &   meV \\
\hline
Mean  & 0.30 & -0.6476  & -0.28 & -4.34 & -0.07\minh \\
Stdev & 0.02 &  0.0018  &  0.02 &  0.05 &  0.05 \\
Max   & 0.32 &  0.6497  &  0.30 &  4.40 &  0.12 \\
RMS   & 0.12 &  0.2644  &  0.12 &  1.77 &  0.03 \\
\end{tabular}
\end{ruledtabular}
\end{table}

To be cautious, we have investigated this issue further by comparing
results of geometry optimizations and frequency calculations on the 12
$(d_{\delta}^Ad_{\delta}^B)$ states of Ni$_2$ (six singlet, broken
symmetry, and six triplet) with B3LYP/AE functional both with and
without charge density fitting.  The results are summarized in
Table~\ref{tab:cdfit-errs}.  Although the errors in total energies are
rather large (on the order of a little less than $1\mHartree$, as can
be seen in column labeled $E_e$/mHartree in
Table~\ref{tab:cdfit-errs}), they all have the same sign, averaging
$-0.6476\pm{}0.0018\mHartree$.  Moreover, the errors in the relative
energies (with respect to the ground state Ni atom, labeled $E_e$/meV
in Table~\ref{tab:cdfit-errs}) are much smaller ($\approx{}5\meV$),
and again all with the same sign.  Finally, the relative ordering of
the states is correct, and the root-mean-square of the relative
energies with respect to the lowest energy state from each
calculation, labeled $\Delta{}E_e$/meV in Table~\ref{tab:cdfit-errs},
is $0.03\meV$ with a maximum of $0.12\meV$.  The maximum error due to
charge density fitting to be expected in exploring the PES's of larger
clusters is on the order of $2$--$3\meV$ per Ni atom.

As stated in Section~\ref{sec:methods}, we did not use charge density
fitting with ECP because of the large errors that resulted when we
tried the use of Ahlrichs Coulomb Fitting basis in combination with
Stuttgart RSC ECP. In Table~\ref{tab:ecp-cdfit-errs} we report the
errors in the bond lengths ($d_e$), dissociation energies ($D_e$), and
harmonic vibrational frequencies ($\omega_e$) of Ni$_2$ and NiH, and
dipole moment ($\mu$) computed with B3LYP and FSLYP functionals using
``Stuttgart RSC ECP'' ECP and basis set with ``Ahlrichs Coulomb
Fitting'' basis. The errors in bond lengths are negligible for both
Ni$_2$ and NiH, but the errors in the vibrational frequencies of both
Ni$_2$ and NiH, diplole moment of NiH and dissociation energy of NiH,
of the order of $5\,\%$, are significat. The error in the dissociation
energy of Ni$_2$ is large, $-0.3\eV$ ($-16\,\%$) for B3LYP and
$-0.33\eV$ ($-25\,\%$) for FSLYP.


\section{Accuracy and convergence issues of DFT computations}\label{sec:Accur:DFT}

%
\begin{table}[tbp]
\caption[Accuracy of the numerical integration and convergence
criteria.]{\label{tab:grids} Details of the integration grid for the
evaluation of the exchange-correlation energy: the number of atomic
radial (rad.) and angular (ang.) shells for Ni and H, along with the
corresponding convergence criteria for the DFT calculations for each
level of accuracy of the numerical integration, in atomic units:
energy ($E$), density ($\rho$) and orbital gradient (orb. grd.).}
\begin{ruledtabular}
\begin{tabular}{lrrrrccc}
              & \multicolumn{2}{c}{Ni} & \multicolumn{2}{c}{H} & \multicolumn{3}{c}{Accuracy}\\
                \cline{2-3}              \cline{4-5}  \cline{6-8}
grid          & rad. & ang. & rad. & ang. & $E$    & $\rho$   & orb. grd.\minh \\
\hline
\verb|xfine|  &    160 &    1454 &    100 &    1202 & $10^{-8}$ & $10^{-7}$ & $10^{-6}$\minh \\
\verb|fine|   &    130 &     974 &     60 &     590 & $10^{-7}$ & $10^{-6}$ & $10^{-5}$ \\
\verb|medium| &    112 &     590 &     45 &     434 & $10^{-6}$ & $10^{-5}$ & $10^{-4}$ \\
\end{tabular}
\end{ruledtabular}
\end{table}

The numerical integration necessary for the evaluation of the
exchange-correlation energy implemented in NWChem uses an
Euler-MacLaurin scheme for the radial components (with a modified
Mura-Knowles transformation) and a Lebedev scheme for the angular
components.  Table~\ref{tab:grids} lists the grid details for the
three levels of accuracy for the numerical integration that are used
in our DFT calculations, labeled by the corresponding keywords from
NWChem ({\verb|medium|}, {\verb|fine|} and {\verb|xfine|}). In the
same table we list convergence criteria used for each level of
accuracy of the numerical integration.

In order to assess the errors arising from numerical integration we
have performed a series of computations using different predefined
grid schemes available in NWChem.  First, we have performed the atomic
calculations using both {\verb|xfine|} and {\verb|fine|} grids.  The
differences are of the order of total energy target accuracy of the
{\verb|fine|} grid ($\approx 1.5\cdot 10^{-7}\Hartree$).  We have also
compared the all-electron DFT computations using B3LYP functional with
{\verb|fine|} grid against the ones with {\verb|xfine|} grid for
geometry optimization and frequency calculations for Ni$_2$,
$(d_{x^2-y^2}^A d_{xy}^B)$ singlet and triplet states. The differences
are on the order of $10^{-4}\uAA$ for equilibrium bond length,
$2\cdot{}10^{-6}\Hartree$ for total equilibrium energy and $0.2\icm$
for vibrational frequency.  We conclude that the {\verb|fine|} grid is
appropriate for geometry optimization and vibrational frequency
calculations, and have used it in the present work.  For the potential
energy curve (PEC) scans we use the {\verb|medium|} grid, which gives
for a 19-point B3LYP/AE PEC scan in the range $2\ldots3.2\uAA$ of Ni$_2$
$(d_{x^2-y^2}^A d_{xy}^B)$ singlet an error in energy of $16\mueV$
(maximum) and $1.4\mueV$ (root-mean-square) with respect to the
{\verb|fine|} grid computations.



\bibliography{309440JCP}

\nocite{Parr:dft_text}
\nocite{Szabo:qc_text}
\nocite{SchaeferIII:el_str.1}
\nocite{SchaeferIII:el_str.2}
\nocite{Parr:el_str}

\end{document}